\documentclass[journal]{new-aiaa} 
\usepackage[utf8]{inputenc}
\usepackage{textcomp}
\usepackage{graphicx}
\usepackage{amsmath}
\usepackage[version=4]{mhchem}
\usepackage{siunitx}
\usepackage{longtable,tabularx}
\usepackage{amsmath,systeme}
\usepackage{cancel}
\usepackage{mathrsfs}
\usepackage{bbm}
\usepackage{caption}
\usepackage{subcaption}
\usepackage{graphicx}
\usepackage{array}
\usepackage{tikz}
\usetikzlibrary{shapes,arrows}
\usepackage{comment}

\graphicspath{{Figs/}}

\newtheorem{rem}{Remark}

\title{Optimal Powered Descent Guidance with \\ Pyramid-Shaped Approach-Angle Constraints}

\author{Revital Frenkel \footnote{Graduate student, The Stephen B. Klein Faculty of Aerospace Engineering. E-mail: revitalfr@campus.technion.ac.il} and Vitaly Shaferman \footnote{Associate Professor, The Stephen B. Klein Faculty of Aerospace Engineering. E-mail: vitalysh@technion.ac.il. (corresponding author)}}
\affil{Technion - Israel Institute of Technology, Haifa 3200003, Israel}

\newcommand{\bfr}{\boldsymbol{r}}

\newcommand{\bfV}{\boldsymbol{V}}

\renewcommand{\eqref}[1]{Eq.~(\ref{#1})}
\newcommand{\eqsref}[1]{Eqs.~(\ref{#1})}
\newcommand{\eqssref}[2]{Eqs.~(\ref{#1})-(\ref{#2})}

\newcommand{\figref}[1]{Fig.~\ref{#1}}

\newcommand{\figssref}[2]{Figs.~\ref{#1}-\ref{#2}}

\newcommand{\secref}[1]{Sec.~\ref{#1}}

\newcommand{\tableref}[1]{Table~\ref{#1}}
\newcommand{\citeref}[1]{Ref.~\cite{#1}}

\renewcommand{\include}{\input}

\begin{document}
\maketitle

\begin{abstract}
In this paper, a novel optimal soft-landing guidance law with inequality approach-angle path constraints is analytically derived. The proposed guidance law prevents ground collision and enables approach-angle control by constraining the optimal trajectory to remain within a convex inverted pyramid originating at the landing point. A 3D point-mass linear kinematic model in a constant gravitational field is employed, together with a quadratic control-effort cost and terminal constraints on position and velocity.
Analytical open-loop and closed-loop solutions, together with the optimal final time, are derived using Pontryagin’s Minimum Principle and the optimality conditions at the transitions between unconstrained and constrained arcs. It is additionally shown that the optimal final time decreases when the path constraints become active. The resulting guidance law is continuous, piecewise linear in time, and nonlinear in the states in closed-loop. When a constraint becomes active, the controller cancels the gravitational component normal to the constraint, causing the trajectory to evolve along the constraint surface.
The proposed guidance law is evaluated in simulations under various initial conditions, demonstrating accurate landing performance and consistent satisfaction of the path constraints.

\end{abstract}

\footnotetext[0]{Presented as Paper 2026-2167 at the AIAA SciTech Forum, Orlando, FL, 12 - 16 January 2026.}

\section*{Nomenclature} \label{Nomenclature}
{\renewcommand\arraystretch{1.0}
\noindent\begin{longtable*}{@{}l @{\quad=\quad} l@{}}
$\mathbf{A}$, $\mathbf{B}$, $\mathbf{C}$  & State, control, and gravity matrices\\
${\bf D}_{nx}$ & Direction cosine matrix from the $(x,y,z)$ frame to the $(n,t,r)$ frame \\
$\mathbf{E}$, $\mathbf{P}$, $\mathbf{Q}$, $\mathbf{W}$ & Switching-condition matrices \\
${f}$, ${\bar f}$, ${\bar{\bar f}}$ & Unconstrained, plane-constrained, and line-constrained optimal-time polynomials, respectively \\
${\bf g}$ & Gravitational acceleration vector, m/s$^2$ \\
$\mathcal{H}$, $\mathcal{G}$, $\mathcal{F}$ & Unconstrained, plane-constrained, and line-constrained Hamiltonians, respectively  \\
$\mathcal{J}$ & Cost function  \\
$L_i$, $M_i$, $N_i$ & Unconstrained, plane-constrained, and line-constrained co-state constants, respectively \\
$m$  & Mass, $kg$ \\
${\bf P}_{nx}$ & State transformation matrix from the $(x,y,z)$ frame to the $(n,t,r)$ frame \\
$\mathbf{r}$   & Position vector, $m$ \\
$S_1$, $S_2$ & Plane and line constraints, respectively\\
$\mathbf{T}$  & Thrust vector, $N$ \\
$t$, $t_f$ & Time and final time, respectively, $s$ \\
$t_1$, $t_2$ & Plane- and line-constraint activation times, respectively, s \\
$t_{go_1}$, $t_{go_2}$, $t_{go_f}$ & Plane-constraint, line-constraint, and final time-to-go values, respectively, s \\
$t_f^*$, $\bar{t}_f^*$, $\bar{\bar{t}}_f^*$ & Unconstrained, plane-constrained, and line-constrained optimal final times, respectively \\
$\mathbf{u}$   & Thrust acceleration vector, $m/s^2$\\
$\mathbf{V}$   & Velocity vector, $m/s$ \\
$\mathbf{x}$  & State vector \\
${\bf y}$, ${\bf z}$ & Plane- and line-constraint tangency-condition vectors, respectively \\
$\boldsymbol{\lambda}$, $\boldsymbol{\mu}$, $\boldsymbol{\eta}$ & Co-states for the unconstrained, plane-constrained, and line-constrained arcs, respectively  \\
$\Delta t_{ij}$ & Time interval from $t_i$ to $t_j$, s \\
${\boldsymbol \Phi}$ & State-transition matrix \\
$\psi$, $\theta$, $\phi$ & 3-2-1 Euler rotation angles from the $(x,y,z)$ frame to the $(n,t,r)$ frame, deg\\
$\left[\mathbf{0}\right]$, $\left[\mathbf{I}\right]$ & Zero and identity matrices, respectively \\
$\mathbbm{1}$ & Unit step function \\
\end{longtable*}} 
\noindent \emph{Superscript}
{\renewcommand\arraystretch{1.0}
\noindent\begin{longtable*}{@{}l @{\quad=\quad} l@{}}
$\Box'$ & Quantity expressed in the $(n,t,r)$ frame \\
$\bar{\Box}'$ & Quantities associated with the plane-constrained reduced dynamics $(t,r)$ \\
$\bar{\bar{\Box}}'$ & Quantities associated with the line-constrained reduced dynamics $(r)$\\
${\Box}^*$ & Optimal quantity\\
${\Box}^-, {\Box}^+$ & Left- and right-hand limits at a switching point, respectively\\
\end{longtable*}} 
\noindent \emph{Subscript}
{\renewcommand\arraystretch{1.0}
\noindent\begin{longtable*}{@{}l @{\quad=\quad} l@{}}
${\Box}_x, {\Box}_y, {\Box}_z$ & Components in the $x$, $y$, and $z$ directions, respectively \\
${\Box}_n, {\Box}_t, {\Box}_r$ & Components in the $n$, $t$, and $r$ directions, respectively \\
${\Box}_0, {\Box}_f$ & Initial and final values, respectively\\
${\Box}_1, {\Box}_2$  & Values at the plane- and line-constraint activation times, respectively, s \\
\end{longtable*}} 

\section{Introduction} \label{Introduction}
Soft-landing guidance, also known as powered-descent guidance, is a fundamental problem in aerospace engineering due to its importance in planetary landing missions, reusable launch vehicles, and Vertical Take-Off and Landing (VTOL) systems. The objective is to guide a vehicle to a prescribed landing site with zero terminal velocity while satisfying mission and safety constraints. In practical scenarios, additional geometric constraints are often required to avoid ground collision and prevent excessively shallow approach angles, making the design of optimal landing guidance laws significantly more challenging.

The Apollo Powered Descent Guidance (APDG) developed in \citeref{APDG} has served as the baseline powered-descent guidance approach for numerous planetary landing studies and missions. APDG assumes a quadratic time-varying thrust-acceleration profile, whose coefficients are determined from the terminal position, velocity, and thrust-acceleration constraints. Another well-known guidance law is E-guidance \cite{Eguidance}, which assumes a linear thrust-acceleration profile. \citeref{lu2019augmented} later introduced the Augmented Apollo Powered Descent Guidance (AAPDG), which generalized both APDG and E-guidance within a unified framework and enabled a family of tunable guidance laws through a single gain parameter.

The guidance laws discussed thus far were not originally derived from optimal-control principles. However, \citeref{1997} later showed that E-guidance minimizes the control effort in a constant gravitational field. 
Optimal-control-based guidance laws are attractive because they can minimize control effort, propellant consumption, flight time, or terminal-state errors while satisfying mission constraints.

Optimal-control problems are commonly solved using either direct or indirect methods. Both approaches can be used to obtain numerical solutions, and a survey of such numerical methods is provided in \citeref{rao2009survey}. Direct methods are generally computationally efficient and relatively insensitive to initialization, enabling convergence even with poor initial guesses. However, the resulting solutions provide limited analytical insight into the structure of the optimal trajectory, while their accuracy and robustness depend on the discretization and optimization techniques employed. Indirect methods, in contrast, derive the necessary optimality conditions and transform the problem into a Two-Point Boundary-Value Problem (TPBVP). However, numerical solution of the resulting TPBVP is generally more sensitive to boundary conditions and initial guesses. Nevertheless, indirect numerical approaches can yield highly accurate solutions and provide deeper insight into the structure of the optimal trajectory.

Indirect methods can also yield analytical guidance laws that are computationally efficient for onboard implementation and provide physical insight into the solution structure and associated trade-offs.
Using an indirect linear-quadratic (LQ) formulation, \citeref{1997} derived an analytical optimal power-descent guidance law with free final time and constrained terminal conditions. Similarly, \citeref{Gutman} employed an LQ formulation with soft terminal constraints imposed through a terminal weighting matrix and a fixed final time.

None of the guidance laws presented thus far explicitly addressed ground collision or low approach angles. This limitation was partially addressed in \citeref{wong2006guidance}, where the problem was divided into two phases by introducing an intermediate point above the landing site using the APDG two-point boundary-value solution \cite{APDG}. However, the resulting guidance law was not optimal. Similar intermediate-point approaches were later proposed in Refs. \cite{OptimalFeedbackIP,ZEMZEV}, where piecewise-optimal LQ solutions were constructed before and after the intermediate point to prevent ground collision. In \citeref{nataf2024optimal}, an optimal-control-based powered-descent guidance law was developed in which the trajectory passes through an optimally selected intermediate point without splitting the optimization problem into two components. 
While these and other intermediate-point approaches enabled ground-collision avoidance and partial control of the terminal approach direction, they did not enforce these requirements through path constraints incorporated directly into the optimal-control formulation.

\citeref{yang2020fuel,you2022theoretical} investigated fuel-optimal powered-descent guidance problems with path constraints. \citeref{yang2020fuel} solved a free-final-time fuel-optimal landing problem with thrust-direction constraints using a convex optimization framework, introducing altitude as the independent variable to facilitate the constraint formulation. In \citeref{you2022theoretical}, a theoretical analysis of a glide-slope-constrained fuel-optimal landing problem was conducted using a state-inequality formulation. It was shown that the optimal thrust magnitude exhibits a bang-bang structure, that the glide-slope constraint can become active only at isolated points, and that the thrust profile may contain more than three subarcs. Additional insights were provided in \citeref{ito2023optimal}, where the constrained fuel-optimal landing problem was solved using an indirect approach. In the presence of thrust-pointing constraints, the solution was shown to exhibit a max–min–max switching structure in the thrust magnitude together with an active–inactive–active constraint-activation sequence. Collectively, these studies demonstrated that path constraints can fundamentally alter the structure of the solution, producing thrust profiles with more than two switching points, unlike the unconstrained case \cite{PropellantOptimal,smoothbangbang}.

However, these and previous studies have shown that fuel-optimal powered-descent guidance laws typically exhibit bang-bang control structures, which may complicate implementation in practical systems. Furthermore, the resulting solutions are generally open-loop, remain saturated near the terminal phase of the descent, and have substantial numerical components, which is often undesirable for practical applications. To mitigate these limitations, \citeref{lu2024rethinking} showed that near-optimal propellant performance can be achieved without bang-bang thrusting by constraining the thrust magnitude to simple continuous, constant, or linear functions of time, thereby simplifying implementation and improving robustness. Similar implementation considerations also motivated the modification of the pure propellant-optimal formulation in \citeref{Lu2026}.

These implementation challenges, together with the difficulty of explicitly enforcing geometric path constraints, motivate the development of alternative optimal-control-based soft-landing guidance formulations. In particular, many existing formulations do not explicitly enforce path constraints required to prevent ground collision or excessively shallow approach angles. Several studies have addressed such constraints, but most existing approaches rely heavily on numerical optimization and frequently yield bang-bang control structures, which may complicate real-time implementation and reduce robustness. Our motivation was therefore to advance analytically based quadratic optimal-control formulations that yield continuous controllers while explicitly incorporating path constraints into the guidance formulation.

\begin{figure}[!ht]
    \centering
    \includegraphics[width=0.2\textwidth]{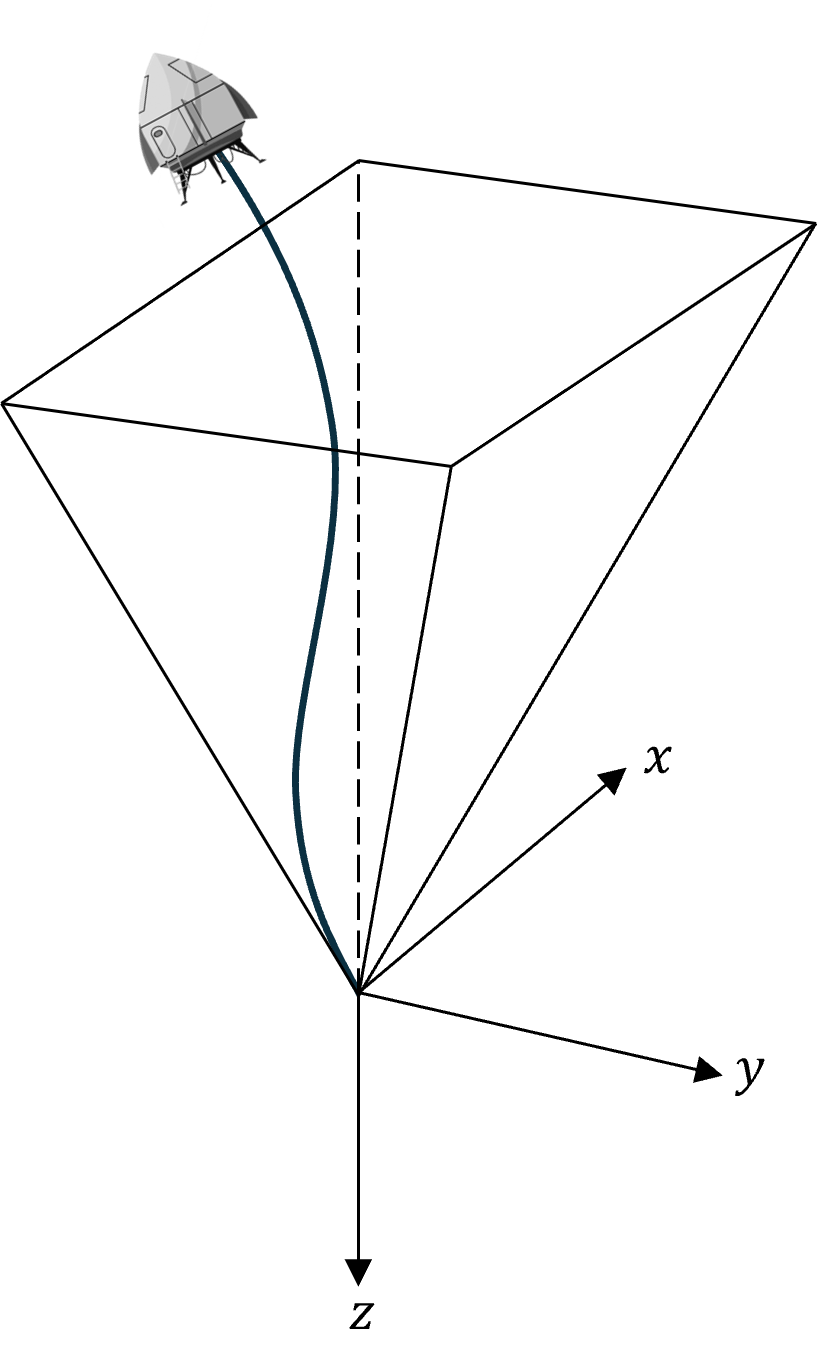}
    \caption{Pyramid path constraint.}
    \label{fig:Pyramid}
\end{figure}

In this paper, we analytically derive a novel optimal-control-based soft-landing guidance law with a quadratic control-effort cost, subject to terminal-state constraints, free final time, linear dynamics, and inequality path constraints, using Pontryagin’s Minimum Principle. The path constraints are modeled using a convex inverted pyramid originating at the landing point, as depicted in \figref{fig:Pyramid}. The derived optimal controller is continuous, piecewise linear in time, and nonlinear in the states in closed loop. When a constraint becomes active, the controller cancels the gravitational component normal to the constraint, causing the trajectory to evolve along the constraint surface. It is also shown that activation of the path constraints reduces the optimal final time. The proposed guidance law is evaluated in simulation under various initial conditions, demonstrating accurate landing trajectories and consistent satisfaction of the pyramid path constraints. An earlier version of the proposed guidance law was presented in \citeref{frenkel2026optimal} and is extended here.

The remainder of the paper is organized as follows. The model derivation and optimization problem formulation are presented in \secref{sec:ModelsDerivation} and \secref{sec:OptimizationProblemFormulation}, respectively. The guidance-law derivation and performance analysis are presented in \secref{sec:GuidanceLawDerivation} and \secref{sec:PerformanceAnalysis}, respectively. Finally, concluding remarks are given in \secref{sec:Conclusions}. Additional derivations are provided in the Appendices.

\section{Models Derivation} \label{sec:ModelsDerivation}

Neglecting the atmospheric drag and under the assumption that the gravitational acceleration is constant, the kinematics, in a non-rotating inertial Cartesian system $(x,y,z)$ attached to the required terminal position on the flat surface of the planet, is
\begin{subequations} \label{eq:3D_motion_eqn}
    \begin{align} \label{r_dot}
      \dot{\bf r} (t)  & =  {\bf V} (t) \\ \label{eq:v_dot}
      \dot{\bf V} (t) & = {\bf u} (t) + \bf g \\            \label{eq:u_from_T} 
      {\bf u} (t) & = {\bf T}(t)/m(t)
    \end{align}
\end{subequations}
where ${{\bf r}= \left[r_x,r_y,r_z \right]^T}$ is the position vector of the lander, ${{\bf V}= \left[V_x,V_y,V_z \right]^T}$ is the velocity vector of the lander, $ {\bf g}  = \left[g_x,g_y,g_z \right]^T $ is the gravitational acceleration, ${\bf u}  = \left[u_x, u_y, u_z \right]^T$ is the thrust acceleration control command and ${\bf T} = \left[T_x,T_y,T_z \right]^T $ denotes the thrust. We define $m(t)$ to be the mass at time $t$. 

\begin{figure} [htbp]
 \centering
    \includegraphics[scale  = 0.2]{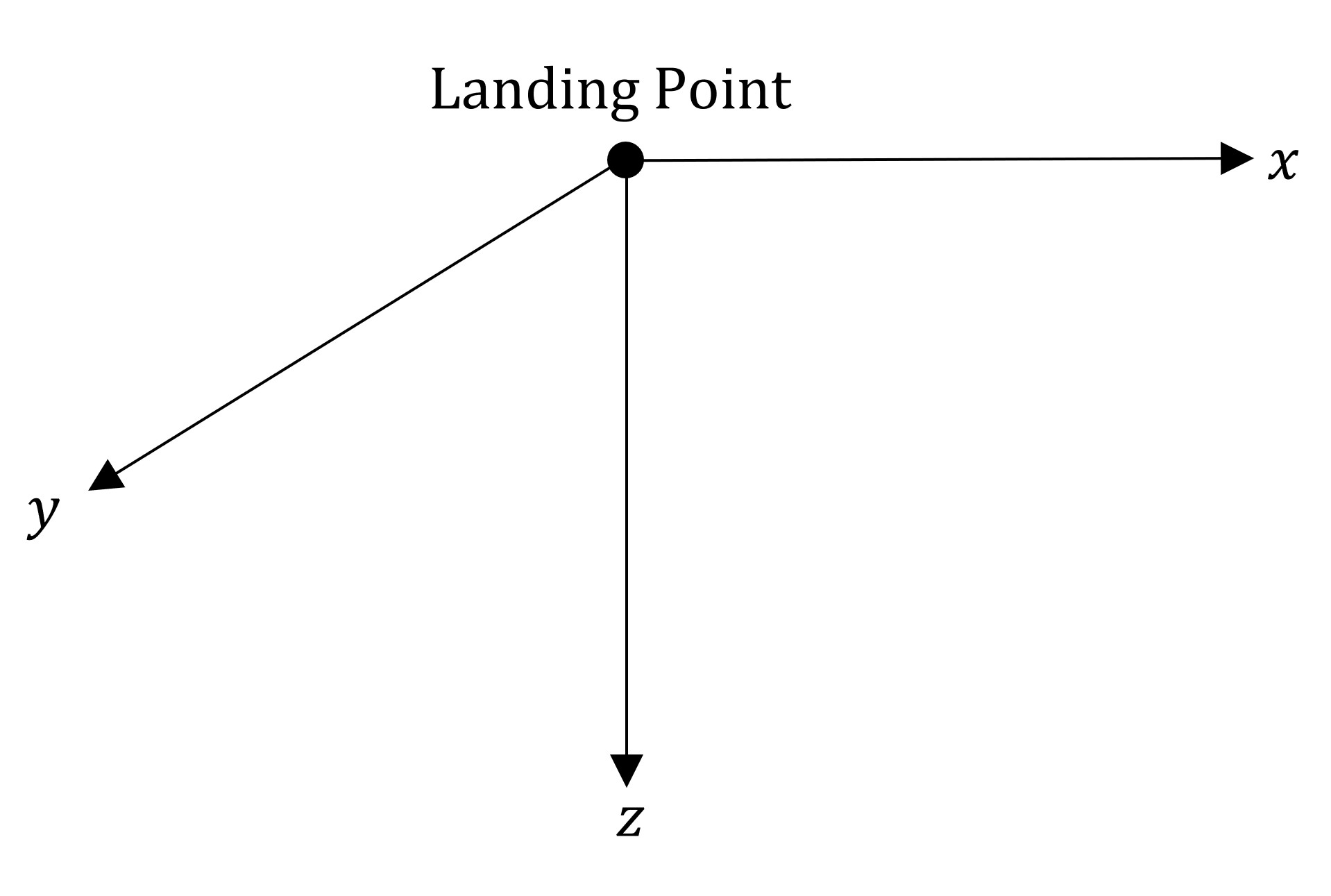}
    \caption{Coordinate system.}
    \label{fig:coordinateSystem}
\end{figure}

The origin of the coordinate system is defined in the landing point with the z-axis pointing downwards. The x-axis is in the down-range direction and the y-axis is in the cross-range direction (see \figref{fig:coordinateSystem}).

The state vector of the problem is 
\begin{equation} \label{eq:3D_state_vec}
{\bf x}  = \begin{bmatrix}
            {\bf r}^T & {\bf V}^T 
        \end{bmatrix}^T
\end{equation}
Thus, 
\begin{equation}
    {\dot{{\bf{x}}}} = \begin{bmatrix}
        {\dot{{\bf{r}}}}^T & {\dot{\bf{V}}}^T 
    \end{bmatrix} ^T  =  \begin{bmatrix}
        {\bf V}^T & {\bf u}^T+{\bf g}^T 
    \end{bmatrix} ^T
\end{equation}
Using the Equations of Motion (EOM) in \eqref{eq:3D_motion_eqn}, we can represent the lander kinematics by
\begin{equation} \label{eq:3D_EOM_xyz}
    {\dot{{\bf{x}}}} = {\bf Ax+Bu+Cg} 
\end{equation}
where
\begin{equation} \label{eq:3D_mat_ABL_3D_full}
    {\bf A} = \begin{bmatrix}
        {\bf{[0]}}_3 & {\bf{[I]}}_3 \\
        {\bf{[0]}}_3 & {\bf{[0]}}_3
    \end{bmatrix}, \quad
    {\bf B} = \begin{bmatrix}
        {\bf{[0]}}_3 \\
        {\bf{[I]}}_3 
    \end{bmatrix}, \quad
    {\bf C} = \begin{bmatrix}
        {\bf{[0]}}_3 \\
        {\bf{[I]}}_3 
    \end{bmatrix} 
\end{equation}
and $\left[\bf{0} \right]_n$ and $ \left[{\bf{I}}\right]_n$ denote a matrix of zeros and the identity matrix with dimensions of $n\times n$, respectively.

The transition matrix associated with \eqref{eq:3D_EOM_xyz}  from time $t$ to $t_*$ is (see Appendix A)
\begin{equation} \label{eq:3D_transition_matrix}
    {\boldsymbol \Phi}(t_*,t) = {\boldsymbol \Phi}(t_{{go_*}}) =  {\begin{bmatrix}
    [{\bf{I}}]_3 & t_{{go_*}}\cdot[{\bf{I}}]_3 \\ 
    [{\bf 0}]_3 & [{\bf{I}}]_3
    \end{bmatrix}}
\end{equation}
where $t_{go_*}=t_*-t$ is the time-to-go till $t_*$. Because the dynamics in \eqref{eq:3D_EOM_xyz} is autonomous, the transition matrix is only a function of a single argument $t_{go_*}$.

\section{Optimization Problem Formulation} \label{sec:OptimizationProblemFormulation}

The cost function we would like to minimize is given by
\begin{equation} \label{eq:3D_cost_func_original}
    \mathcal{J} = 
    \frac{1}{2} \int_{0}^{t_f} {{\Vert {\bf u}\Vert }^2} \ dt
\end{equation}
with the following terminal constraints
\begin{equation} \label{eq:terminal_conditions}
    {\bf{x}}_f = {\bf{x}}(t_f) ={\bf{{0}}}
\end{equation}

\begin{figure}[ht!]
    \centering
    \includegraphics[width=0.25\textwidth]{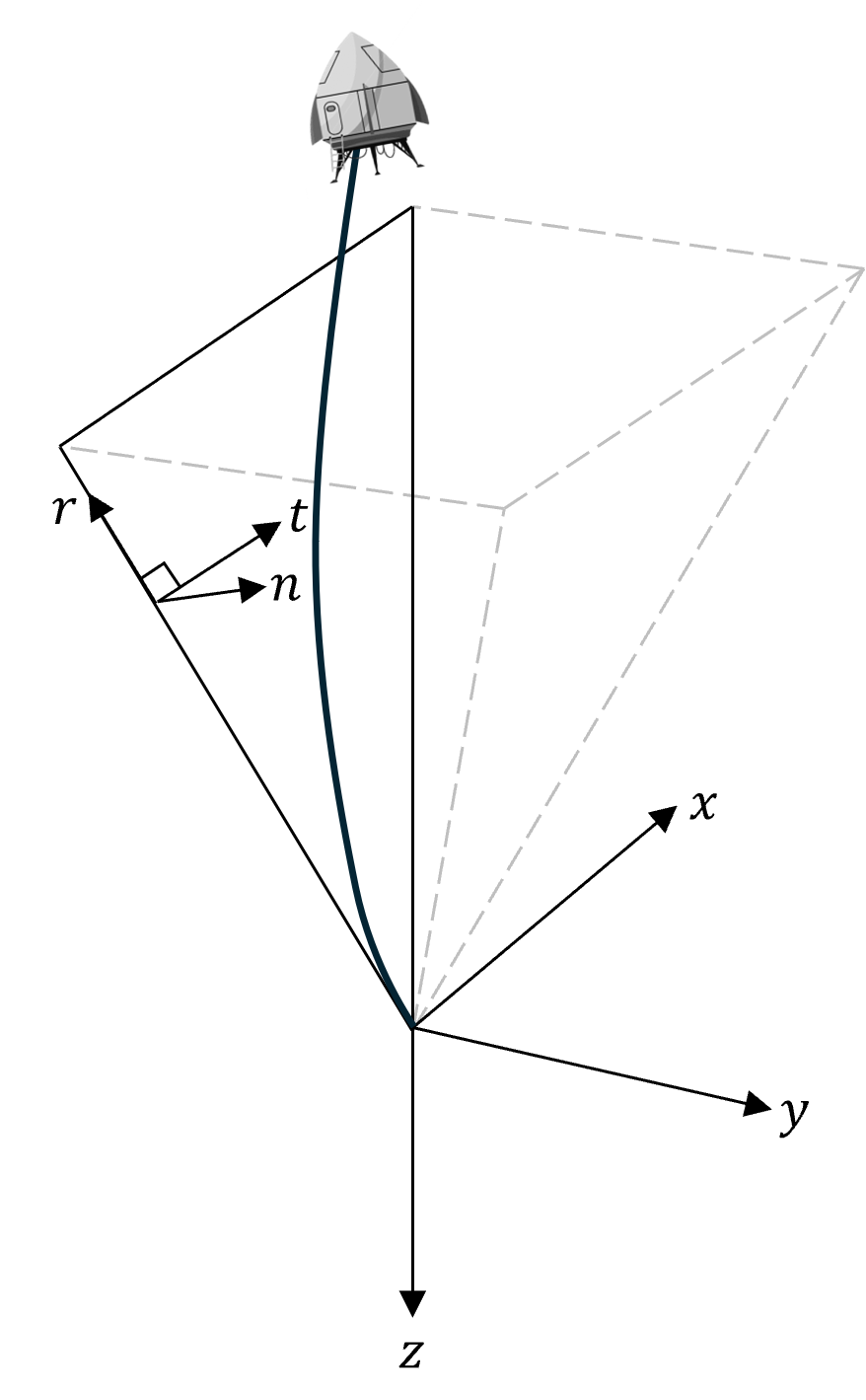}
    \caption{Triangular face constraint demonstration and coordinate rotation.}
    \label{fig:rotationDef}
\end{figure}

We consider additional path constraints in the form of an inverted pyramid. The pyramid is composed of a convex base, in which the initial position lies, and triangular faces, as depicted in \figref{fig:rotationDef}. Without loss of generality, the derivation will concentrate on a single triangular face, but the algorithm will be applied to the full pyramid. Each triangular face lies in a plane, and the triangle's edges lie on lines. To simplify the derivation, we introduce a new inertial Cartesian coordinate system $(n, t, r)$, where $n$ is normal to the plane constraint, $t$ is in the plane normal to the line constraint, and $r$ completes a right-handed frame. Both $n$ and $t$ are oriented inward, pointing into or along the pyramid's interior convex hull. This coordinate system is obtained from the original $(x, y, z)$  frame by a standard 3-2-1 Euler sequence with Euler angles, $\psi,\ \theta$, and $\phi$, respectively. The first two angles $\psi$ and $\theta$ rotate the $x$ axis into alignment with $n$, and the final rotation $\phi$ is applied about $n$ to align the original $y$ axis with $t$, perpendicular to the line constraint pointing into the constrained triangle (see \figref{fig:rotationDef}).

Note that when solving for a plane constraint, the rotation angle $\phi$ can be chosen arbitrarily, but when solving a full pyramid constraint problem, $\phi$ is determined from the geometry of every two intersecting planes (see Appendix B for the derivation of $\phi$ in the case of a square pyramid).

The rotation matrix from the original $(x, y, z)$ frame to the $(n, t, r)$ frame is
\begin{equation}
        {\bf D}_{nx} = \begin{bmatrix}
            \cos\theta\cos\psi & \cos\theta\sin\psi & -\sin\theta \\
            \sin\phi\sin\theta\cos\psi -\cos\phi\sin\psi & \cos\phi\cos\psi + \sin\phi\sin\theta\sin\psi & \sin\phi\cos\theta \\
            \sin\phi\sin\psi + \cos\phi\sin\theta\cos\psi & \cos\phi\sin\theta\sin\psi -\sin\phi\cos\psi & \cos\phi\cos\theta
        \end{bmatrix} 
\end{equation}
and the inverse transformation is given by ${\bf D}_{xn}={\bf D}_{nx}^{-1}={\bf D}_{nx}^{T}$.
The state vector represented in $(n,t,r)$ frame is
\begin{equation} \label{eq:3D_state_vec_ntr}
    {\bf x}' = {\bf P}_{nx}{\bf x} = \begin{bmatrix}
            {\bf r'}^T & {\bf V'}^T 
        \end{bmatrix}^T, \quad {\bf r'} = \begin{bmatrix}
            r_n & r_t & r_r
        \end{bmatrix}^T, \quad {\bf V'} = \begin{bmatrix}
            V_n & V_t & V_r 
        \end{bmatrix}^T
\end{equation}
where the coordinate transformations for the full state vector are 
\begin{equation} \label{eq:Pnx_Pxn} 
        \boldsymbol{P}_{nx} = \begin{bmatrix}
            {\bf D}_{nx} & [\boldsymbol{0}]_3 \\
            [\boldsymbol{0}]_3 & {\bf D}_{nx}
        \end{bmatrix} , \quad \boldsymbol{P}_{xn} = \boldsymbol{P}_{nx}^{-1} = \boldsymbol{P}_{nx}^{T}
        = \begin{bmatrix}
            {\bf D}_{xn} & [\boldsymbol{0}]_3 \\
            [\boldsymbol{0}]_3 & {\bf D}_{xn}
        \end{bmatrix}
\end{equation}
and the control and the gravitation vectors in the $(n,t,r)$ frame are
\begin{equation} \label{eq:3D_ug_vec_ntr} 
    {\bf u'} = {\bf D}_{nx}{\bf u} = \begin{bmatrix}
            u_n & u_t & u_r
        \end{bmatrix}^T, \quad {\bf g'} = {\bf D}_{nx}{\bf g} = \begin{bmatrix}
            g_n & g_t & g_r 
        \end{bmatrix}^T
\end{equation}

When defining the constraints, two phases must be considered. First, to ensure the trajectory remains above the defined plane, the constraint formulation requires the component normal to the plane to remain non-negative at all times. Second, to remain inside the triangle, the same geometric principle is applied using the direction normal to the line (in the plane). These constraints can be written in both $(n,t,r)$ and $(x,y,z)$ frames, respectively 
\begin{equation} \label{eq:3D_constraint_nt} 
        S({\bf{x}'},t) = \begin{bmatrix}
            r_n \geq 0 \\
            r_t \geq 0
        \end{bmatrix} 
\end{equation}
and
\begin{equation}  \label{eq:3D_constraint_xy} 
        S({\bf{x}},t) = \begin{bmatrix}
            r_x \cos\theta\cos\psi + r_y \cos\theta\sin\psi -r_z \sin\theta \geq 0 \\
            r_x \left(\sin\phi\sin\theta\cos\psi - \cos\phi\sin\psi \right) + r_y \left(\cos\phi\cos\psi + \sin\phi\sin\theta\sin\psi \right) +  r_z \left(\sin\phi\cos\theta\right) \geq 0 \\
        \end{bmatrix}
\end{equation}

Since the constraints formulation in the $(n, t, r)$ frame (\eqref{eq:3D_constraint_nt}) is more straightforward and offers a more intuitive geometric interpretation, we choose to solve the problem in the constraint-aligned frame $(n, t, r)$ rather than in the original $(x, y, z)$ frame.

Using the EOM in  \eqref{eq:3D_EOM_xyz} and \eqsref{eq:3D_state_vec_ntr} and (\ref {eq:3D_ug_vec_ntr}), we can represent the kinematics in $(n,t,r)$ frame by
\begin{equation} \label{eq:3D_EOM_ntr}
    {\dot{{\bf{x}}}}' = {\bf P}_{nx}{\dot{{\bf{x}}}} =  {\bf P}_{nx}{\bf AP}_{nx}^{-1}{\bf x}'+{\bf P}_{nx}{\bf B}{\bf D}_{nx}^{-1}{\bf u}'+{\bf P}_{nx}{\bf C}{\bf D}_{nx}^{-1}{\bf g'} = {\bf Ax'+Bu'+Cg'}
\end{equation}
Therefore, the transition matrix in the $(n, t, r)$ frame is identical to the one in the $(x, y, z)$
\begin{equation}
    {\boldsymbol \Phi}'(t_{go_*}) = {\boldsymbol \Phi}(t_{go_*})
\end{equation}

The cost function and terminal constraints, in $(n,t,r)$ coordinates are
\begin{equation} \label{eq:3D_cost_fun_ntr}
    \mathcal{J}' = 
    \frac{1}{2} \int_{0}^{t_f} {{\Vert {\bf u}'\Vert }^2} \ dt, \quad
    {\bf x}'_f = {\bf{0}}
\end{equation}

\section{Guidance Law Derivation} \label{sec:GuidanceLawDerivation}

The main idea of the derivation is that when the unconstrained trajectory intersects the path constraint, the optimal solution transitions from an unconstrained arc to a constrained arc that remains on the active constraint.
This assumption will be proved optimal in \secref{subsec:ProofofOptimality} by showing that, once the trajectory reaches the constraint, application of the unconstrained optimal controller causes it to evolve along the constraint rather than depart from it.
As described in \secref{sec:OptimizationProblemFormulation}, the guidance law derivation will concentrate on a single triangular face of the pyramid constraint, which defines a plane. If the trajectory violates one of the pyramid planes, it transitions onto the corresponding triangular face, thus entering a constrained arc. Should the trajectory further violate the boundary of a triangle, it then transitions onto an edge, i.e., from a constrained trajectory on a plane to one on a line. 
Let us assume an unconstrained arc for $t < t_1$, followed by a constrained arc for $t_1 \leq t < t_2$ where the trajectory is restricted to lie on a plane, and finally a constrained arc for $t_2 \leq t \leq t_f$ where the trajectory is further restricted to lie on a line. In any other scenario, only part of these transitions will occur, and only a partial part of the derivation will have to be solved. To improve readability, these partial, simpler derivations are presented in Appendices C and D. 

Let us define: 
\begin{equation} \label{eq:tgo12f}
    t_{go_1} = t_1 - t ,\quad t_{go_2} = t_2 - t ,\quad t_{go_f} = t_f - t ,\quad \Delta t_{12} = t_2 - t_1 = t_{go_2} - t_{go_1} ,\quad \Delta t_{2f} = t_f - t_2 = t_{go_f} - t_{go_2}
\end{equation}

\subsection{Unconstrained Arc Solution}
The Hamiltonian of the problem is
\begin{equation} \label{eq:3D_H}
    \begin{split}
        \mathcal{H} =\lambda_1 V_n + \lambda_2 V_t + \lambda_3 V_r &+ \lambda_4 (u_n + g_n) + \lambda_5 (u_t + g_t) + \lambda_6 (u_r + g_r) + \frac{1}{2}\left(u_n^{2} + u_t^{2} + u_r^{2}\right), \\ \boldsymbol{\lambda}^T &= \begin{bmatrix}
        \lambda_1 & \lambda_2 & \lambda_3 & \lambda_4 & \lambda_5 & \lambda_6
    \end{bmatrix}
    \end{split}
\end{equation}
The adjoint equations are
\begin{equation}\label{eq:3D_lambda_dot}
    \dot{\boldsymbol{\lambda}}^T(t)= -\dfrac{\partial \mathcal{H}}{\partial {\bf{x}}'} = -\begin{bmatrix}
        0 & 0 & 0 & \lambda_1 & \lambda_2 & \lambda_3
    \end{bmatrix}  
\end{equation}
Integrating \eqref{eq:3D_lambda_dot},  where $L_i,\ i\in\{1,\dots,6\}$ are the co-states at $t_1$, yields
\begin{equation} \label{eq:3D_lambda}
     \boldsymbol{\lambda}^T(t_{go_1}) = \begin{bmatrix}
       L_1 & L_2 & L_3 & \left(L_1\cdot t_{go_1} + L_4\right) & \left(L_2\cdot t_{go_1} + L_5\right) & \left(L_3\cdot t_{go_1} + L_6\right)
   \end{bmatrix}, \quad t\in [0,t_1]
\end{equation}

The optimal controller satisfies
\begin{equation}
    \frac{\partial \mathcal{H}}{\partial {\bf{u}'}} = {\bf{0}}^T = \boldsymbol{\lambda}^T{\bf{B}} + {\bf{u}'}^T 
\end{equation}
\begin{equation} \label{eq:3D_u_unconstrained}
    \Rightarrow\quad {\bf{u}'}^* = -{\bf{B}}^T \boldsymbol{\lambda} = -\begin{bmatrix}
        \lambda_4 & \lambda_5 & \lambda_6
    \end{bmatrix}^T = -\begin{bmatrix}
        \left(L_1\cdot t_{go_1} + L_4\right) & \left(L_2\cdot t_{go_1} + L_5\right) & \left(L_3\cdot t_{go_1} + L_6\right)
    \end{bmatrix}^T
\end{equation}

Note that if the optimal trajectory does not violate a plane constraint at any point during the scenario, the boundary condition at $t_f$ can be used to find $L_{1-6}$ (see Appendix C for the solution). In this case $t_{go_1}=t_{go_f}$.

\subsection{Plane-Constrained Arc} \label{sec:PlaneConstrainedArc}
\subsubsection{Reduced-Order Dynamics on a Plane-Constrained Arc} \label{sec:ReducedDynamicsDerivation}

The plane state inequality constraint in the $(n,t,r)$ coordinates is
\begin{equation}
    S_1({\bf{x}'},t) = r_n \geq 0
\end{equation}
Let the tangency conditions, ${\bf{y}}({\bf{x}'},t)$, be defined as
\begin{equation}
    {\bf{y}}({\bf{x}'},t) = \begin{bmatrix}
        S_1({\bf{x}'},t) \\
        S_1^{(1)}({\bf{x}'},t) \\
        \vdots \\
        S_1^{(p-1)}({\bf{x}'},t)
    \end{bmatrix} \quad,\quad S_1^{(n)} = \frac{d^{(n)}S_1}{dt^{(n)}}
\end{equation}
and let $C_1({\bf{x}'},{\bf{u}'},t)$ be defined as the first derivative of the constraint in which the control vector explicitly appears, where $p$ is the constraint order
\begin{equation} 
    C_1({\bf{x}'},{\bf{u}'},t) = S_1^{(p)}({\bf{x}'},{\bf{u}'},t) 
\end{equation}
On a constrained arc, the constraint and its derivatives up to the $p$ derivative must be null. On the plane constraint, we get a second-order constraint, and
\begin{equation} \label{C_1}
    {\bf{y}}({\bf{x}'},t) = \begin{bmatrix}
        r_n \\
        V_n
    \end{bmatrix} = \boldsymbol{0},\quad C_1({\bf{x}'},{\bf{u}'},t) = \dot{V}_n=u_n + g_n = 0,\quad t \in [t_1,t_f]
\end{equation}
Since the dynamics are 6-dimensional, and the constraint is of second order, the dynamics in \eqref{eq:3D_EOM_ntr} are reduced to the following 4-dimensional plane-constrained reduced states
\begin{equation} \label{eq:3D_state_vec_reduced}
    \bar{{\bf{x}}}'  = 
    \begin{bmatrix}
        {\bar{\bf r}}^{'T} & {\bar{\bf V}}^{'T} 
        \end{bmatrix}^T, \quad {\bar{\bf r}'} = \begin{bmatrix}
            r_t & r_r
        \end{bmatrix}^T, \quad {\bar{\bf V}'} = \begin{bmatrix}
            V_t & V_r 
        \end{bmatrix}^T
\end{equation}
with the following EOM
\begin{equation} \label{eq:3D_EOM_reduced1}
    {\dot{\bar{{\bf{x}}}}}' =  \begin{bmatrix}
    {\dot{r}_t} & {\dot{r}_r} & {\dot{V}_t} & {\dot{V}_r} 
    \end{bmatrix} ^T  =  \begin{bmatrix}
        V_t & V_r & u_t+g_t & u_r+g_r
    \end{bmatrix} ^T
\end{equation}
Using the EOM in  \eqref{eq:3D_EOM_reduced1}, we can represent the lander kinematics on the plane constraint by
\begin{equation} \label{eq:3D_dynamics_reduced}
    {\dot{\bar{{\bf{x}}}}}' = {\bar{\bf{A}}\bar{\bf{x}}'}+{\bar{\bf{B}}}{\bar{\bf{u}}}'+\bar{\bf{C}}{\bar{\bf{g}}}'
\end{equation}
where
\begin{equation} \label{eq:3D_mat_ABL_3D_reduced1}
    {\bf \bar{A}} = \begin{bmatrix}
        {\bf{[0]}}_2 & {\bf{[I]}}_2 \\
        {\bf{[0]}}_2 & {\bf{[0]}}_2
    \end{bmatrix}, \quad
    {\bf \bar{B}} = \begin{bmatrix}
        {\bf{[0]}}_2 \\
        {\bf{[I]}}_2 
    \end{bmatrix}, \quad
    {\bf \bar{C}} = \begin{bmatrix}
        {\bf{[0]}}_2 \\
        {\bf{[I]}}_2 
    \end{bmatrix} , \quad
    {\bf \bar{g}}' = \begin{bmatrix}
        g_t \\
        g_r 
    \end{bmatrix}, \quad 
    {\bf \bar{u}}' = \begin{bmatrix}
        u_t \\
        u_r
    \end{bmatrix} , \quad 
    \bar{{\boldsymbol \Phi}}'(t_{go_*}) = {\begin{bmatrix}
    [{\bf{I}}]_2 & t_{{go_*}}\cdot[{\bf{I}}]_2 \\ 
    [{\bf 0}]_2 & [{\bf{I}}]_2
    \end{bmatrix}}
\end{equation}
From this point onward, the notation $\bar{\Box }$ refers to the reduced dynamics on the plane constraint.

\subsubsection{Plane Constrained Arc Solution}

From \eqref{C_1}, we can get the optimal controller normal to the plane constraint
\begin{equation} \label{eq:3D_un}
    u_n + g_n = 0 \quad\Rightarrow\quad u_n^* = -g_n ,\quad t\in\left[ t_1,t_f \right]
\end{equation}
To find the optimal controller tangential to the plane constraint, let us define the constrained Hamiltonian based on \eqref{eq:3D_EOM_reduced1}, the cost function in \eqref{eq:3D_cost_fun_ntr}, and \eqref{eq:3D_un}, and a new reduced set of co-states $\boldsymbol{\mu}$
\begin{equation} \label{eq:3D_G}
    \mathcal{G} = \mu_1 V_t + \mu_2 V_r + \mu_3(u_t+g_t) + \mu_4(u_r+g_r) +\frac{1}{2}\left( g_n^2 + u_t^2 + u_r^2 \right),\quad \boldsymbol{\mu}^T = \begin{bmatrix}
        \mu_1 & \mu_2 & \mu_3 & \mu_4
    \end{bmatrix}
\end{equation}
The plane-constrained adjoint equations are
\begin{equation}\label{eq:3D_mu_dot}
    \dot{\boldsymbol{\mu}}^T(t)= -\dfrac{\partial \mathcal{G}}{\partial {\bar{\bf{x}}}'} = -\begin{bmatrix}
        0 & 0 & \mu_1 & \mu_2
    \end{bmatrix}  
\end{equation}
Integrating \eqref{eq:3D_mu_dot}, where $M_i,\ i\in\{1,2,3,4\}$ are the plane-constrained co-states at $t_2$, yields
\begin{equation} \label{eq:3D_mu}
  \boldsymbol{\mu}^T(t_{go_2}) = \begin{bmatrix}
       M_1 & M_2 & (M_1\cdot t_{go_2} + M_3) & (M_2\cdot t_{go_2} + M_4) 
   \end{bmatrix}, \quad t\in [t_1,t_2] 
\end{equation}

The optimal controller tangential to the plane constraint satisfies
\begin{equation} \label{eq:3D_u_constrained}
    \frac{\partial \mathcal{G}}{\partial {\bar{{\bf{u}}}}'} = {\bf{0}}^T 
    \quad\Rightarrow\quad \bar{{\bf{u}}}'^* = -\begin{bmatrix}
        \mu_3 & \mu_4
    \end{bmatrix}^T = -\begin{bmatrix}
        (M_1\cdot t_{go_2} + M_3) & (M_2\cdot t_{go_2} + M_4)
    \end{bmatrix}^T
\end{equation}

Note that if the optimal trajectory does not violate the line constraint at any point during the scenario, the boundary condition can be used to find $M_{1-4}$ (see Appendix D for the solution). In this case $t_{go_2}=t_{go_f}$.

\subsection{Line-Constrained Arc} \label{sec:LineConstrainedArc}
\subsubsection{Reduced-Order Dynamics on a Line-Constrained Arc} \label{sec:ReducedDynamicsDerivation}
In addition to the active plane constraint, the line state inequality constraint in the $(n,t,r)$ coordinates is
\begin{equation}
    S_2({\bar{\bf{x}}}',t) = r_t \geq 0
\end{equation}
Let the tangency conditions ${\bf{z}}({\bar{\bf{x}}}',t)$ and $C_2({\bar{\bf{x}}}',{\bar{\bf{u}}}',t)$ be defined as
\begin{equation}
    {\bf{z}}({\bar{\bf{x}}}',t) = \begin{bmatrix}
        S_2({\bar{\bf{x}}}',t) \\
        S_2^{(1)}({\bar{\bf{x}}}',t) \\
        \vdots \\
        S_2^{(p-1)}({\bar{\bf{x}}}',t)
    \end{bmatrix} \quad,\quad S_2^{(n)} = \frac{d^{(n)}S_2}{dt^{(n)}}
\end{equation}
\begin{equation}
    C_2({\bar{\bf{x}}}',{\bar{\bf{u}}}',t) = S_2^{(p)}({\bf{x}'},{\bf{u}'},t) 
\end{equation}
On a constrained arc, the constraint and its derivatives up to the $p$ derivative must be null. On the line constraint, we get an additional second-order constraint
\begin{equation} \label{C_2}
    {\bf{z}}({\bar{\bf{x}}}',t) = \begin{bmatrix}
        r_t \\
        V_t
    \end{bmatrix} = \boldsymbol{0},\quad C_2({\bar{\bf{x}}}',{\bar{\bf{u}}}',t) = \dot{V}_t=u_t + g_t = 0,\quad t \in [t_2,t_f]
\end{equation}
Since the plane-constrained dynamics are 4-dimensional, and the line constraint is of second order, the dynamics in \eqref{eq:3D_dynamics_reduced} are reduced again to the following 2-dimensional line-constrained reduced states and EOM
\begin{equation} \label{reduces_state_vec}
    \bar{\bar{{\bf{x}}}}'  = \begin{bmatrix}
        r_r & V_r 
    \end{bmatrix}^T
\end{equation}
\begin{equation} \label{reduces_EOM}
    {\dot{\bar{\bar{{\bf{x}}}}}}' =  \begin{bmatrix}
    {\dot{r}_r} & {\dot{V}_r} 
    \end{bmatrix} ^T  =  \begin{bmatrix}
        V_r & u_r+g_r 
    \end{bmatrix} ^T
\end{equation}
Using the EOM in \eqref{reduces_EOM}, we can represent the lander kinematics on the line constraint by
\begin{equation}
    {\dot{\bar{\bar{{\bf{x}}}}}}' = {\bar{\bar{\bf{A}}}\bar{\bar{\bf{x}}}'}+{\bar{\bar{\bf{B}}}}u_r+{\bar{\bar{\bf{C}}}} g_r
\end{equation}
where
\begin{equation} \label{mat_ABL_2D}
    {\bf \bar{\bar{A}}} = \begin{bmatrix}
    0 &1 \\
    0 & 0
    \end{bmatrix}, \quad
    {\bf \bar{\bar{B}}} = \begin{bmatrix}
    0 \\
    1 
    \end{bmatrix}, \quad
    {\bf \bar{\bar{C}}} = \begin{bmatrix}
    0 \\
    1 
    \end{bmatrix}  , \quad 
    \bar{\bar{{\boldsymbol \Phi}}}'(t_{go_*}) = {\begin{bmatrix}
    1 & t_{{go_*}} \\ 
    0 & 1
    \end{bmatrix}}
\end{equation}
From this point onward, the notation $\bar{\bar{\Box }}$ refers to the reduced dynamics on the line constraint.

\subsubsection{Line-Constrained Arc Solution} \label{subsec:LineConstrainedArcSolution}

From \eqref{C_2} we can get the optimal controller normal to the line constraint
\begin{equation} \label{eq:2D_ut}
    u_t + g_t = 0 \quad\Rightarrow\quad u_t^* = -g_t ,\quad t\in\left[ t_2,t_f \right]
\end{equation}
To find the optimal controller tangential to the constraint, let us define the constrained Hamiltonian based on Eqs. (\ref{reduces_EOM}), (\ref{eq:3D_cost_fun_ntr}), (\ref{eq:3D_un}), and (\ref{eq:2D_ut})
\begin{equation} \label{eq:2D_F}
    \mathcal{F} = \eta_1 V_r + \eta_2 \left( u_r + g_r\right) + \frac{1}{2}\left( g_n^2 + g_t^2 + u_r^2 \right), \quad \boldsymbol{\eta}^T = \begin{bmatrix}
        \eta_1 & \eta_2
    \end{bmatrix}
\end{equation}
The reduced set of co-states $\boldsymbol{\eta}$ is given by
\begin{equation} \label{eq:2D_eta}
    \dot{\boldsymbol{\eta}}^T(t)= -\dfrac{\partial \mathcal{F}}{\partial \bar{\bar{{\bf{x}}}}'} = \begin{bmatrix}
        0 & -\eta_1
    \end{bmatrix} \quad\Rightarrow\quad \boldsymbol{\eta}^T(t_{go_f}) = \begin{bmatrix}
    N_1 & (N_1\cdot t_{go_f} + N_2) \end{bmatrix}, \quad t\in [t_2,t_f] 
\end{equation}
where $N_i,\ i\in\{1,2\}$ are the line-constrained co-states at $t_f$.

The optimal controller tangential to the line constraint yields
\begin{equation} \label{eq:2D_ur_temp}
    \frac{\partial \mathcal{F}}{\partial u_r} = 0 \quad\Rightarrow\quad \eta_2 + u_r = 0 \quad\Rightarrow\quad u_r^* = -\eta_2=-(N_1\cdot t_{go_f} + N_2),\quad t\in\left[ t_2,t_f \right]
\end{equation}
Using the transition matrix, the dynamics, the optimal controller, and the terminal constraints, we can find $\boldsymbol{\eta}(t_{go_f})$ explicitly
\begin{equation}
    {\bf{\bar{\bar{\bf{x}}}}}'_f  = \bar{\bar{\boldsymbol{\Phi}}}'(t_{go_f}) {\bf\bar{\bar{{\bf{x}}}}}'(t) + \int_{t}^{t_f}{\bar{\bar{\boldsymbol{\Phi}}}'(t_f-\tau)\bar{\bar{{\bf{B}}}}u_r^*(\tau)d\tau} + \int_{t}^{t_f}{\bar{\bar{\boldsymbol{\Phi}}}'(t_f-\tau)\bar{\bar{{\bf{C}}}}{g_r}d\tau} = {\bf{0}}
\end{equation}
\begin{equation}  \label{eq:N1_N2}
    \Rightarrow\quad \begin{bmatrix}
        N_1 \\
        N_2
    \end{bmatrix} = \frac{12}{t_{go_f}^4} \begin{bmatrix}
        t_{go_f}r_r(t) + \dfrac{t_{go_f}^2}{2}V_r(t)\\
        -\dfrac{t_{go_f}^2}{2}r_r(t) - \dfrac{t_{go_f}^3}{6}V_r(t) + \dfrac{t_{go_f}^4}{12}g_r
    \end{bmatrix}
\end{equation}
Substituting \eqref{eq:N1_N2}  into \eqref{eq:2D_ur_temp} yields the following optimal controller
\begin{equation} \label{eq:2D_ur}
    u_r^* = -\frac{6r_r(t)}{t_{go_f}^2} - \frac{4V_r(t)}{t_{go_f}} - g_r ,\quad t\in\left[ t_2,t_f \right]
\end{equation}
Note that this controller has the same structure as the unconstrained optimal controller in Appendix C and in \citeref{1997}.

\subsection{Switching Points Conditions}

At the switch points between the constrained and unconstrained arcs (i.e., $t=t_1$ and $t=t_2$), the following conditions must be satisfied \cite{speyer1968optimal} (where $t_i^-$ and $t_i^+$ denote the limits from the left and from the right, respectively, as $t \to t_i$)
\begin{subequations} \label{eq:3D_junc_cond}
    \begin{align}
    \boldsymbol{\mu}^T(t_1^+) &= \boldsymbol{\lambda}^T(t_1^-)\bf{E} \label{eq:3D_junc_cond_1}  \\
    \mathcal{G}(t_1^+) &= \mathcal{H}(t_1^-) + \boldsymbol{\lambda}^T(t_1^-) \left( {\bf{W}}{\bf{y}}_t + {\bf{E}}{\bar{{\bf{x}}}}'_t \right) \label{eq:3D_junc_cond_2} \\
    \boldsymbol{\eta}^T(t_2^+) &= \boldsymbol{\mu}^T(t_2^-)\bf{Q} \label{eq:2D_junc_cond_1}  \\
    \mathcal{F}(t_2^+) &= \mathcal{G}(t_2^-) + \boldsymbol{\mu}^T(t_2^-) \left( {\bf{P}}{\bf{z}}_t + {\bf{Q}}\bar{\bar{{\bf{x}}}}'_t \right) \label{eq:2D_junc_cond_2}
    \end{align}
\end{subequations}
where $\bf{W}$, $\bf{E}$, $\bf{P}$, and $\bf{Q}$ are given by 
\begin{equation} \label{eq:MN_PQ_formula}
    \begin{bmatrix}
        \bf{W} & : & \bf{E}
    \end{bmatrix} = \begin{bmatrix}
        {\bf{y}}_{\bf{x}'} \\ 
        \cdot \cdot \\
        \bar{{\bf{x}}}'_{\bf{x}'}
    \end{bmatrix}^{-1} ,\quad 
    \begin{bmatrix}
        \bf{P} & : & \bf{Q}
    \end{bmatrix} = \begin{bmatrix}
        {\bf{z}}_{\bar{\bf{x}}'} \\ 
        \cdot \cdot \\
        \bar{\bar{{\bf{x}}}}'_{\bar{\bf{x}}'}
    \end{bmatrix}^{-1} 
\end{equation}
and
\begin{equation}
    {\bf{y}}_{{\bf{x}}'} = \dfrac{\partial {\bf{y}}}{\partial {\bf{x}}'} ,\quad \bar{{\bf{x}}}'_{\bf{x}'} = \dfrac{\partial \bar{{\bf{x}}}'}{\partial {\bf{x}}'} ,\quad {\bf{z}}_{\bar{\bf{x}}'} = \dfrac{\partial {\bf{z}}}{\partial {\bar{\bf{x}}}'} ,\quad \bar{\bar{{\bf{x}}}}'_{\bar{\bf{x}}'} = \dfrac{\partial \bar{\bar{{\bf{x}}}}'}{\partial {\bar{\bf{x}}}'} ,\quad {\bf{y}}_t = \dfrac{\partial {\bf{y}}}{\partial t}  ,\quad \bar{{\bf{x}}}'_t = \dfrac{\partial \bar{{\bf{x}}}'}{\partial t} ,\quad {\bf{z}}_t = \dfrac{\partial {\bf{z}}}{\partial t}  ,\quad \bar{\bar{{\bf{x}}}}'_t = \dfrac{\partial \bar{\bar{{\bf{x}}}}'}{\partial t}
\end{equation}
Which yield 
\begin{equation} \label{eq:Time_partial_derivs}
   {\bf{y}}_t = {\bf 0},\quad \bar{{\bf{x}}}'_t = {\bf 0}, \quad
   {\bf{z}}_t = {\bf 0},\quad \bar{\bar{{\bf{x}}}}'_t = {\bf 0}
\end{equation}
and
\begin{equation}\label{eq:MN_PQ}
\mathbf{W}=
\begin{bmatrix}
1 & 0 & 0 & 0 & 0 & 0 \\
0 & 0 & 0 & 1 & 0 & 0
\end{bmatrix}^{T},
\; \; \;
\mathbf{E}=
\begin{bmatrix}
0 & 1 & 0 & 0 & 0 & 0 \\
0 & 0 & 1 & 0 & 0 & 0 \\
0 & 0 & 0 & 0 & 1 & 0 \\
0 & 0 & 0 & 0 & 0 & 1
\end{bmatrix}^{T},
\; \; \;
\mathbf{P}=
\begin{bmatrix}
1 & 0 & 0 & 0 \\
0 & 0 & 1 & 0
\end{bmatrix}^{T},
\; \; \;
\mathbf{Q}=
\begin{bmatrix}
0 & 1 & 0 & 0 \\
0 & 0 & 0 & 1
\end{bmatrix}^T
\end{equation}
Using the switching-point conditions in \eqref{eq:3D_junc_cond} and the tangency conditions in \eqref{C_1} and \eqref{C_2}, the costate integration constants $L_i$ and $M_j$, together with the switching times $t_1$ and $t_2$, can be obtained. The solution proceeds sequentially, starting from the second switching point.

\subsubsection{Second Switching Point $t=t_2$}
Substituting \eqsref{eq:3D_mu}, (\ref{eq:2D_eta}), and (\ref{eq:MN_PQ}) into \eqref{eq:2D_junc_cond_1} yields
\begin{equation} \label{eq:2D_costates_cont}
    \begin{cases}
        \eta_1(t_2) = \mu_2(t_2) = M_2 = N_1 \\
        \eta_2(t_2) = \mu_4(t_2) = M_4 = N_1 \Delta t_{2f} + N_2
    \end{cases}
\end{equation}
Substituting \eqsref{eq:3D_G}, (\ref{eq:2D_F}), and (\ref{eq:Time_partial_derivs}) into \eqref{eq:2D_junc_cond_2} (where $ \Box ^+,\ \Box ^-$ marks $\Box (t_2^+),\ \Box (t_2^-) $ respectively) , yields
\begin{equation}
\begin{split}
   \eta_1^+ V_r &+ \eta_2^+ (u_r^+ + g_r) + \frac{1}{2}\left[g_n^{2} + g_t^{2} + (u_r^+)^{2}\right] \\
   &= \mu_1^- V_t + \mu_2^- V_r + \mu_3^- (u_t^- + g_t) + \mu_4^- (u_r^- + g_r) + \frac{1}{2}\left[g_n^{2} + (u_t^-)^{2} + (u_r^-)^{2}\right] 
   \end{split}
\end{equation}
and substituting 
 \eqref{C_2}, \eqref{eq:2D_ut}, and \eqref{eq:2D_costates_cont} yields for $t=t_2$
\begin{equation} \label{eq:2D_Hamiltonian_cont}
    -\mu_3^- \left( g_t + u_t^- \right) + \mu_4^- \left( u_r^+ - u_r^- \right) = \frac{1}{2} \left[(u_t^-)^{2} - g_t^{2}\right] + \frac{1}{2} \left[(u_r^-)^{2} - (u_r^+)^{2}\right]
\end{equation}
Substituting the optimal controllers in \eqref{eq:3D_u_constrained} and  \eqref{eq:2D_ur_temp} and \eqref{eq:2D_costates_cont} we can find $M_3$
\begin{equation} \label{eq:2D_M3}
    M_3 \left( M_3-g_t  \right) = \frac{1}{2} \left(M_3^{2} - g_t^{2}\right) \quad\Rightarrow\quad M_3 = g_t
\end{equation}
Substituting the optimal controller in \eqref{eq:3D_u_constrained} into the dynamics in \eqref{eq:3D_dynamics_reduced} and integrating, we can find the plane-constrained reduced states at $t=t_2$
\begin{equation}
    \bar{{\bf{x}}}'_2 = \bar{{\bf{x}}}'(t_2) = \bar{{\boldsymbol{\Phi}}}'(t_{go_2}) \bar{{\bf{x}}}'(t) + \int_t^{t_2}{\bar{{\boldsymbol{\Phi}}}'(t_2-\tau)\bar{\bf{B}}\bar{{\bf{u}}}'^*(\tau)d\tau} + \int_t^{t_2}{\bar{{\boldsymbol{\Phi}}}'(t_2-\tau)\bar{\bf{C}}{\bar{\bf{g}}}'d\tau}
\end{equation}
\begin{equation} \label{eq:2D_x2}
    \Rightarrow\quad \bar{{\bf{x}}}'_2 = \begin{bmatrix}
    [{\bf I}]_2 & t_{{go_2}}\cdot[{\bf I}]_2 \\ 
    [{\bf 0}]_2 & [{\bf I}]_2
    \end{bmatrix} \begin{bmatrix}
        {\bf r}'(t) \\
        {\bf V}'(t)
    \end{bmatrix} - \begin{bmatrix}
        \frac{M_1}{3}t_{go_2}^3 + \frac{M_3}{2}t_{go_2}^2 \\
        \frac{M_2}{3}t_{go_2}^3 + \frac{M_4}{2}t_{go_2}^2 \\
        \frac{M_1}{2}t_{go_2}^2 + M_3t_{go_2} \\
        \frac{M_2}{2}t_{go_2}^2 + M_4t_{go_2}
    \end{bmatrix} + \begin{bmatrix}
        \frac{t_{go_2}^2}{2}[{\bf I}]_2 \\
        t_{go_2}[{\bf I}]_2
    \end{bmatrix}\bar{{\bf{g}}}'
\end{equation}
where $\bar{{\bf{x}}}'_2$ is the plane-constrained reduced state vector at $t = t_2$.

To determine $M_1$, Eqs. (\ref{C_2}) and (\ref{eq:2D_M3}) can be substituted into the third component of \eqref{eq:2D_x2} yielding
\begin{equation} \label{eq:2D_M1}
    V_t(t_2) = V_t(t) - \frac{M_1}{2}t_{go_2}^2 - g_t t_{go_2} + g_t t_{go_2}= 0 \quad\Rightarrow\quad M_1 = \frac{2V_t(t)}{t_{go_2}^2}
\end{equation}

To determine $t_{go_2}$, Eqs. (\ref{C_2}), (\ref{eq:2D_M3}), and (\ref{eq:2D_M1}) can be substituted into the first component of \eqref{eq:2D_x2} yielding
\begin{equation}\label{eq:tgo2}
    r_t(t_2) = r_t(t) + V_t(t) t_{go_2} - \frac{2V_t(t)}{3t_{go_2}^2}t_{go_2}^3 - \frac{g_t}{2}t_{go_2}^2 + \frac{g_t}{2} t_{go_2}^2= 0 \quad\Rightarrow\quad t_{go_2} = -\frac{3r_t(t)}{V_t(t)}
\end{equation}
%

\subsubsection{First Switching Point $t=t_1$} \label{subsec:First_Switching_Point}
Substituting \eqsref{eq:3D_lambda}, (\ref{eq:3D_mu}), and (\ref{eq:MN_PQ}) into \eqref{eq:3D_junc_cond_1} and then substituting \eqsref{eq:2D_costates_cont}, (\ref{eq:2D_M3}), and (\ref{eq:2D_M1}) yields
\begin{equation} \label{eq:3D_costates_cont}
    \begin{cases}
        \mu_1(t_1) = \lambda_2(t_1) = L_2 = M_1 = \frac{2V_t(t)}{t_{go_2}^2} \\
        \mu_2(t_1) = \lambda_3(t_1) = L_3 = M_2 = N_1 \\
        \mu_3(t_1) = \lambda_5(t_1) = L_5 = M_1\cdot \Delta t_{12} + M_3 = \frac{2V_t(t)}{t_{go_2}^2}\Delta t_{12} + g_t \\
        \mu_4(t_1) = \lambda_6(t_1) = L_6 = M_2\cdot \Delta t_{12} + M_4 = N_1\cdot \Delta t_{12} + N_1 \Delta t_{2f} + N_2  
    \end{cases}
\end{equation}
Substituting \eqsref{eq:3D_H}, (\ref{eq:3D_G}), and (\ref{eq:Time_partial_derivs}) into \eqref{eq:3D_junc_cond_2} (where $\Box ^+,\ \Box ^-$ marks $\Box (t_1^+),\ \Box (t_1^-)$, respectively), yields 
\begin{equation}
    \begin{split} 
      \mu_1^+ V_t &+ \mu_2^+ V_r + \mu_3^+ (u_t^+ + g_t) + \mu_4^+ (u_r^+ + g_r) + \frac{1}{2}\left[g_n^{2} + (u_t^+)^{2} + (u_r^+)^{2}\right] = \\
      &= \lambda_1^- V_n + \lambda_2^- V_t + \lambda_3^- V_r + \lambda_4^- (u_n^- + g_n) + \lambda_5^- (u_t^- + g_t) + \lambda_6^- (u_r^- + g_r) + \frac{1}{2}\left[(u_n^-)^{2} + (u_t^-)^{2} + (u_r^-)^{2}\right]
\end{split}
\end{equation}
and substituting \eqref{C_1}, \eqref{eq:3D_un}, and \eqref{eq:3D_costates_cont} yields for $t=t_1$
\begin{equation} \label{eq:3D_Hamiltonian_cont}
    -\lambda_4^- \left( g_n + u_n^- \right) + \lambda_5^- \left( u_t^+ - u_t^- \right) + \lambda_6^- \left( u_r^+ - u_r^- \right)  = \frac{1}{2} \left[(u_n^-)^{2} - g_n^{2}\right] + \frac{1}{2} \left[(u_t^-)^{2} - (u_t^+)^{2}\right] + \frac{1}{2} \left[(u_r^-)^{2} - (u_r^+)^{2}\right]
\end{equation}
Substituting the optimal controllers in Eqs. (\ref{eq:3D_u_unconstrained}) and (\ref{eq:3D_u_constrained}), and \eqref{eq:3D_costates_cont} we can find $L_4$
\begin{equation} \label{eq:3D_L4}
    L_4 \left( L_4-g_n  \right) = \frac{1}{2} \left(L_4^{2} - g_n^{2}\right) \quad\Rightarrow\quad L_4 = g_n
\end{equation}

Substituting the optimal controller in \eqref{eq:3D_u_unconstrained} into the dynamics in \eqref{eq:3D_EOM_ntr} and integrating, we can find the position and velocity at $t=t_1$
\begin{equation}
    {\bf{x}}'_1 = {\bf{x}}'(t_1) = {\boldsymbol{\Phi}'}(t_{go_1}) {\bf{x}}'(t) + \int_t^{t_1}{{\boldsymbol{\Phi}}'(t_1-\tau)\bf{B}{\bf{u}}'^*(\tau)d\tau} + \int_t^{t_1}{{\boldsymbol{\Phi}}'(t_1-\tau)\bf{C}{\bf{g}}'d\tau}
\end{equation}
\begin{equation} \label{eq:3D_x1}
    \Rightarrow\quad {\bf{x}}'_1 = \begin{bmatrix}
    [{\bf I}]_3 & t_{{go_1}}\cdot[{\bf I}]_3 \\ 
    [{\bf 0}]_3 & [{\bf I}]_3
    \end{bmatrix} \begin{bmatrix}
        {\bfr}'(t) \\
        {\bfV}'(t)
    \end{bmatrix} - 
    \begin{bmatrix}
        \dfrac{L_1}{3}t_{go_1}^3 + \dfrac{L_4}{2}t_{go_1}^2 \\
        \dfrac{L_2}{3}t_{go_1}^3 + \dfrac{L_5}{2}t_{go_1}^2 \\
        \dfrac{L_3}{3}t_{go_1}^3 + \dfrac{L_6}{2}t_{go_1}^2 \\
        \dfrac{L_1}{2}t_{go_1}^2 + L_4t_{go_1} \\
        \dfrac{L_2}{2}t_{go_1}^2 + L_5t_{go_1} \\
        \dfrac{L_3}{2}t_{go_1}^2 + L_6t_{go_1}
    \end{bmatrix} + 
    \begin{bmatrix}
        \dfrac{t_{go_1}^2}{2}[{\bf I}]_3 \\
        t_{go_1}[{\bf I}]_3
    \end{bmatrix}{\bf{g}}'
\end{equation}
where ${\bf{x}}'_1$ is the state vector at $t = t_1$.

To determine $L_1$, Eqs. (\ref{C_1}) and (\ref{eq:3D_L4}) can be substituted into the fourth component of \eqref{eq:3D_x1} yielding
\begin{equation} \label{eq:3D_L1}
    V_n(t_1) = V_n(t) - \frac{L_1}{2}t_{go_1}^2 - g_nt_{go_1} +g_n  t_{go_1}= 0 \quad\Rightarrow\quad L_1 = \frac{2V_n(t)}{t_{go_1}^2}
\end{equation}

To determine $t_{go_1}$, Eqs. (\ref{C_1}), (\ref{eq:3D_L4}), and (\ref{eq:3D_L1}) can be substituted into the first component of \eqref{eq:3D_x1} yielding
\begin{equation} \label{eq:3D_tgo1}
    r_n(t_1) = r_n(t) +V_n(t)  t_{go_1}- \frac{2V_n(t)}{3t_{go_1}^2}t_{go_1}^3 - \frac{g_n}{2}t_{go_1}^2 + \frac{g_n}{2}t_{go_1}^2 = 0 \quad\Rightarrow\quad t_{go_1} = -\frac{3r_n(t)}{V_n(t)}
\end{equation}
%
\subsection{The Combined Guidance Law} \label{subsec:TheCombinedGuidanceLaw}
Combining the controllers associated with the three phases yields
%
\begin{equation} \label{eq:3D_optimal_controller_temp} 
    {\bf{u}'}^* = \begin{cases}
        -\begin{bmatrix}
             L_1\cdot t_{go_1} + L_4 & L_2\cdot t_{go_1} + L_5 & L_3\cdot t_{go_1} + L_6
        \end{bmatrix}^T & ,t\leq t_1 \\
        -\begin{bmatrix}
             g_n &
             M_1\cdot t_{go_2} + M_3 & M_2\cdot t_{go_2} + M_4 
        \end{bmatrix}^T & ,t_1 < t \leq t_2 \\
        -\begin{bmatrix}
             g_n &
             g_t &
             N_1 t_{go_f} + N_2
        \end{bmatrix}^T & ,t> t_2
    \end{cases}
\end{equation}
Substituting Eqs. (\ref {eq:2D_costates_cont}), (\ref{eq:2D_M3}) (\ref{eq:3D_costates_cont}), (\ref{eq:3D_L4}), and (\ref{eq:tgo12f}) yields
\begin{equation} \label{eq:3D_optimal_controller_temp_smooth} 
    {\bf{u}'}^* = \begin{cases}
        -\begin{bmatrix}
             L_1\cdot t_{go_1} + g_n & M_1\cdot t_{go_2} + g_t &  N_1 t_{go_f} + N_2
        \end{bmatrix}^T & ,t\leq t_1 \\
        -\begin{bmatrix}
             g_n &
             M_1\cdot t_{go_2} + g_t &  N_1 t_{go_f} + N_2 
        \end{bmatrix}^T & ,t_1 < t \leq t_2 \\
        -\begin{bmatrix}
             g_n &
             g_t &
             N_1 t_{go_f} + N_2
        \end{bmatrix}^T & ,t> t_2
    \end{cases}
\end{equation}
Note that the controller is continuous at the switching points ($t_{go_1}=0$ and $t_{go_2}=0$). Furthermore, \eqref{eq:3D_optimal_controller_temp_smooth} shows that the optimal controller is piecewise linear in time. The $n$-component, associated with the plane-constrained arc, undergoes a slope change at $t_1$, while the $t$-component, associated with the line-constrained arc, undergoes a slope change at $t_2$. The $r$-component, however, remains linear throughout and exhibits no slope changes. In addition, whenever a constraint becomes active, the controller exactly cancels the gravitational component normal to the active constraint.

Substituting Eqs. (\ref{eq:N1_N2}), (\ref{eq:2D_M1}), (\ref{eq:tgo2}), (\ref{eq:3D_L1}), and (\ref{eq:3D_tgo1}) we can write the optimal controller as a function of the current state, using the unit step function
\begin{equation} \label{eq:3D_optimal_controller} 
    {\bf{u}}'^* = -\begin{bmatrix}
        g_n -\dfrac{2V_n^2(t)}{3r_n(t)}\cdot  \mathbbm{1}(t_1-t) \\
        g_t -\dfrac{2V_t^2(t)}{3r_t(t)} \cdot \mathbbm{1}(t_2-t) \\
        g_r + \dfrac{6r_r(t)}{t_{go_f}^2} + \dfrac{4V_r(t)}{t_{go_f}}
    \end{bmatrix}, \quad \mathbbm{1}(\tau) = \begin{cases}
        0 ,& \tau < 0 \\
        1 ,& \tau \geq 0
    \end{cases}
\end{equation}

This is a closed-loop controller that uses ${\bf r}'(t)$ and ${\bf V}'(t)$ to compute the optimal control input ${\bf u}'^*(t)$. Furthermore, the controller is nonlinear in the $n$- and $t$-directions, which are normal to the constraints, and linear in the direction of the line constraint. On the constrained arcs, no closed-loop action is applied in the direction normal to the active constraint. This follows from the assumption that the tangency conditions are satisfied exactly at the switching points, thereby eliminating the need for feedback in that direction.

In practice, however, the tangency conditions may be violated slightly due to numerical errors, modeling uncertainties, or external disturbances. Such deviations can propagate along the constrained arc and lead to terminal position and velocity errors. To mitigate these effects and improve satisfaction of the terminal constraints, an alternative optimal closed-loop controller can be introduced in the direction normal to the active constraint. This controller actively corrects deviations from the constraint and minimizes the resulting terminal errors. The corresponding formulation is discussed in Remark~\ref{rem:Closedloop_On_The_Constraint} in the next section (\secref{subsec:ProofofOptimality}).
%

\subsection{Optimality of Remaining on the Constraint} \label{subsec:ProofofOptimality}

To prove that the optimal trajectory does not leave a constrained arc once it reaches it, we will show that the optimal unconstrained solution keeps the trajectory on the constraint when initiated on the constraint.

The following optimal unconstrained solution is derived in Appendix C in \eqref{eq:optimal_u_unconstrained}
\begin{equation} \label{eq:optimal_u_unconstrained_2}
{\bf u}'^* = -\frac{6{\bf{r}}'(t)}{t_{go_f}^2} - \frac{4{\bf{V}}'(t)}{t_{go_f}} - {\bf{g}}'
\end{equation}
When the trajectory reaches the constraint, regardless of whether it remains on it or not, it must be tangent to the constraint. Therefore, ${\bf{y}}({\bf{x}'},t) = \boldsymbol{0}$ in \eqref{C_1} must hold on the plane constraint and additionally ${\bf{z}}({\bar{\bf{x}}}',t) =\boldsymbol{0}$ in \eqref{C_2} must hold on the line constraint. In other words, both the position and velocity components normal to the constraint must vanish. Without loss of generality, we consider $r_n = V_n = 0$.
Substituting $r_n$ and $V_n$ into \eqref{eq:optimal_u_unconstrained_2} yields
\begin{equation}
    u_n^* = -g_n
\end{equation}
which is identical to the optimal controller in \eqref{eq:3D_optimal_controller} and therefore ensures that the optimal trajectory remains on the constraint. This proves that the trajectory cannot depart from the constraint, since the unconstrained and constrained optimal controllers are identical when the constraint is active.
\begin{rem} \label{rem:Closedloop_On_The_Constraint}
Since the unconstrained closed-loop optimal controller in the direction normal to the constraint was shown to keep the optimal trajectory on the constraint, it may be used along the constrained arcs to provide a closed-loop controller that optimally compensates for possible deviations. It is actually the optimal minimum effort correction to the constraint at the terminal time. 
\end{rem}

\subsection{Optimal and Admissible Final Times} \label{subsec:FinalTime}
\subsubsection{Optimal Final Time} \label{subsec:OptimalFinalTime}

In addition to the optimal controller, the optimal final time can also be derived.
To determine the optimal final time, the following condition on the Hamiltonian $\mathcal{F}(\bar{\bar{t}}_f^{\ *})$ at the final time, arising from the transversality condition, must be satisfied
\begin{equation}  \label{eq:tranversality}
    \mathcal{F}(\bar{\bar{t}}_f^{\ *})=0
\end{equation}
Where $\bar{\bar{t}}_f^{\ *}$ denotes the final optimal time if both the plane and the line constraints become active. 
Since the dynamical system is autonomous and the time does not appear explicitly in the cost, the Hamiltonians $\mathcal{H}$ in \eqref{eq:3D_H}, $\mathcal{G}$ in \eqref{eq:3D_G}, and $\mathcal{F}$ in \eqref{eq:2D_F} are constant along the optimal trajectory \cite{BenAsher}. 
Combining the condition in \eqref{eq:tranversality} and the switching point conditions in \eqref{eq:3D_junc_cond_2} and \eqref{eq:2D_junc_cond_2} (which state that the Hamiltonians are continuous, due to the autonomy of the system and because the constraints do not explicitly include the time), we get the condition
\begin{equation} \label{eq:optimal_time_eqn_temp}
    \mathcal{H}(t) = \mathcal{H}_n(t) + \mathcal{H}_t(t) + \mathcal{H}_r(t) = 0
\end{equation}
that can be used to find the optimal final time, where
\begin{subequations} \label{eq:H_n_t_r}
    \begin{align}
        \mathcal{H}_n(t) &= \lambda_1 V_n + \lambda_4 (u_n^* + g_n) +  \frac{1}{2}\left(u_n^*\right)^{2} \label{eq:H_n}\\
        \mathcal{H}_t(t) &= \lambda_2 V_t + \lambda_5 (u_t^* + g_t) + \frac{1}{2}\left(u_t^*\right)^{2} \label{eq:H_t}\\
        \mathcal{H}_r(t) &= \lambda_3 V_r + \lambda_6 (u_r^* + g_r) + \frac{1}{2}\left(u_r^*\right)^{2}\label{eq:H_r}
    \end{align}    
\end{subequations}
Note that equation \eqref{eq:optimal_time_eqn_temp} holds on the optimal trajectory, and we do not carry the star mark (i.e., $*$) in all the variables (besides the controller) to simplify the notations. Substituting Eqs. (\ref{eq:3D_costates_cont}), (\ref{eq:3D_L4}), (\ref{eq:3D_L1}), and (\ref{eq:3D_optimal_controller}) into \eqref{eq:H_n_t_r} we can derive \eqssref{eq:H_n}{eq:H_t} explicitly, and rewrite \eqref{eq:H_r} similarly to \cite{1997} (which derived the optimal final time in the unconstrained case)
\begin{subequations}\label{eq:Hn_Ht_Hr_2const}
    \begin{align}
        \mathcal{H}_n(t) &= \frac{1}{2}g_n^{2} \\
        \mathcal{H}_t(t) &= \frac{1}{2}g_t^{2} \\
        \mathcal{H}_r(t) &= -\frac{18r_r^2}{t_{go_f}^4} -\frac{12r_r V_r}{t_{go_f}^3} -\frac{2V_r^2}{t_{go_f}^2} + \frac{1}{2}g_r^{2}
    \end{align}    
\end{subequations}
Substituting \eqref{eq:Hn_Ht_Hr_2const} with the initial conditions at $t_0$ into \eqref{eq:optimal_time_eqn_temp} and rearranging yields the following 4-th order polynomial for $\bar{\bar{t}}_f^{\ *}$
\begin{equation} \label{eq:find_optimal_tf}
   {{\Vert {\bf g}'\Vert }^2}\left(\bar{\bar{t}}_f^{\ *}\right)^4 - 4V_r^2(t_0)\left(\bar{\bar{t}}_f^{\ *}\right)^2 - 24r_r(t_0)V_r(t_0)\bar{\bar{t}}_f^{\ *} - 36r_r^2(t_0) = 0
\end{equation}
which can be used to determine the optimal final time.
Note that the solution depends only on the initial conditions in the $r$ axis and the norm of the whole gravitational vector.
This result has a similar structure to the unconstrained case, which appeared in \cite{1997} and is derived in Appendix C.

Another key result is that the optimal final time decreases when the constraints become active
\begin{equation}
    0 \leq \bar{\bar{t}}_f^{\ *} \leq \bar{t}_f^{\ *} \leq t_f^{\ *}
\end{equation}
where $\bar{\bar{t}}_f^{\ *}$, $\bar{t}_f^{\ *}$, and $t_f^{\ *}$ denote that both the plane and the line constraints become active, only the plane constraint becomes active, and the unconstrained cases, respectively (see Appendix E for the full derivation).

\subsubsection{Admissible Final Times} \label{LimitationsoftheOptimalFinalTime}
We now determine the admissible final times required to ensure ground-collision avoidance along the line-constrained arc. To this end, we analyze the initial position and velocity components along the $r$-axis. Unlike the analyses in Refs. \cite{ZEMZEV,wang2021two}, the rotated coordinate system considered here introduces additional combinations of initial conditions that do not arise in the formulations studied previously. As a result, the feasibility analysis becomes more involved.

Substituting the conditions at $t=0$ into \eqref{eq:N1_N2}  yields
\begin{equation} \label{eq:N1N2_Appx_E}
        N_1 = \frac{12r_{r0}}{ t_{f}^3} + \dfrac{6V_{r0}}{ t_{f}^2}, \quad 
        N_2=-\dfrac{6r_{r0}}{ t_{f}^2} - \dfrac{2V_{r0}}{ t_{f}} + g_r, \quad r_r(t=0) \triangleq r_{r0}, \quad V_r(t=0) \triangleq V_{r0}
    \end{equation}
Note that this substitution is valid at the initial time because the $r$-axis controller and kinematics are identical in all three phases of the solution.
Substituting \eqref{eq:N1N2_Appx_E} into \eqref {eq:2D_ur_temp} and substituting the result into the EOM in \eqref{reduces_EOM} yields
\begin{equation} \label{eq:Vr_dot_Appx_E}
        \dot{V}_r (t) =  u_r+g_r = -\left(\frac{12r_{r0}}{t_{f}^3} + \dfrac{6V_{r0}}{ t_{f}^2}\right) \left(t_f-t\right) +\left(\dfrac{6r_{r0}}{t_{f}^2} + \dfrac{2V_{r0}}{t_{f}}\right) \\
    \end{equation}
integrating with the final conditions in \eqref{eq:3D_cost_fun_ntr} yields
\begin{subequations}\label{eq:Vr_rr}
    \begin{align} \label{eq:Vr_Appx_E}
        V_r(t) &= \left(\frac{12r_{r0}}{t_{f}^3} + \frac{6V_{r0}}{t_{f}^2}\right)\cdot \frac{(t_f-t)^2}{2} - \left(\frac{6r_{r0}}
        {t_{f}^2} + \frac{2V_{r0}}{t_{f}}\right)\cdot (t-t_f) \\ 
        r_r(t) &= -\left(\frac{12r_{r0}}{t_{f}^3} + \frac{6V_{r0}}{t_{f}^2}\right)\cdot \frac{(t_f-t)^3}{6} + \left(\frac{6r_{r0} }{t_{f}^2}+ \frac{2V_{r0}}{t_{f}}\right)\cdot \frac{(t_f-t)^2}{2} \label{eq:rr_Appx_E}
    \end{align}
\end{subequations}
The flight-time limits can be determined by analyzing the roots of \eqref{eq:Vr_rr}. We consider each combination of the initial position and velocity signs separately. 
Note that the $r$-axis may point either toward or away from the triangular constraint face. In fact, for each face, one edge is associated with an inward-pointing $r$-axis, whereas the other is associated with an outward-pointing $r$-axis. For brevity we will only analyze the inward-pointing $r$-axis and present the final result for the outward-pointing case.

For an inward-pointing $r$-axis, admissible trajectories must approach the terminal conditions from the positive-$r$ side. Therefore, $r_r(t_f)=0$ must be a local minimum of $r_r(t)$, which requires the following condition:
\begin{equation}
    \dot{V}_r\left(t=t_f\right)\geq 0 \quad \Rightarrow \quad\left(\dfrac{6r_{r0}}{t_{f}^2} + \dfrac{2V_{r0}}{t_{f}}\right) \geq 0
\end{equation}
This yields the following conditions on the final time as a function of the signs of $r_{r0}$ and $V_{r0}$:
\begin{enumerate}
    \item $r_{r0}>0\ ,V_{r0}>0$,
    \begin{equation}
        -\frac{3r_{r0}}{V_{r0}} < 0 <t_{f}
    \end{equation}
    Owing to the signs of $r_{r0}$ and $V_{r0}$, the resulting inequality is satisfied for all positive values. This case is usually omitted from the literature because it imposes no restriction on the flight time and corresponds to a trajectory whose initial velocity is directed away from the landing site.

    \item $r_{r0}>0\ ,V_{r0}<0$,
        \begin{equation}
            0 < t_{f} \leq -\frac{3r_{r0}}{V_{r0}}
        \end{equation}
        This corresponds to the standard condition commonly considered in the literature, where the constraint is imposed along the $z$-axis.

    \item $r_{r0}<0\ ,V_{r0}>0$,
    \begin{equation}
        -\frac{3r_{r0}}{V_{r0}} \leq  t_{f} 
    \end{equation}
This is a novel case in which the lander starts on the negative side of the $r$-axis, passes above the landing site, and approaches the terminal conditions from the positive-$r$ side at touchdown.

    \item $r_{r0}<0\ ,V_{r0}<0$,

    This case admits no feasible solution with an approach from the positive-$r$ side. However, activation of the considered plane constraint is impossible, as the trajectory would encounter the opposite constraint plane first. Therefore, this case can be excluded from the analysis.
\end{enumerate}

For an outward-pointing $r$-axis, admissible trajectories must approach the terminal conditions from the negative-$r$ side.
Therefore, $r_r(t_f)=0$ must be a local maximum of $r_r(t)$, which requires the following condition:
\begin{equation}
    \dot{V}_r\left(t=t_f\right)\leq 0 \quad \Rightarrow \quad\left(\dfrac{6r_{r0}}{t_{f}^2} + \dfrac{2V_{r0}}{t_{f}}\right) \leq 0
\end{equation}
These conditions exactly complement the previous results (i.e., the opposite results in each case) and are therefore not repeated.

If the optimal final time computed from \eqref{eq:find_optimal_tf} lies outside the admissible range for a given case, it is replaced by the nearest admissible value to prevent ground collision.


\section{Performance Analysis} \label{sec:PerformanceAnalysis}
\subsection{Sample Run}
The simulation scenario is based on the lunar landing case presented in \cite{1997}, with modified initial conditions chosen to demonstrate the path constraints incorporated in the proposed guidance law. The initial conditions are ${\bf r}_0 = \left[152400,\ 30480,\ -15240\right]^T~(\mathrm{m})$,
${\bf V}_0 = \left[-914.4,\ 0,\ 200\right]^T~(\mathrm{m/s})$, and
${\bf g} = \left[0,\ 0,\ g_z\right]^T$, where $g_z = 1.625~(\mathrm{m/s^2})$.
The final time is free and is computed from \eqref{eq:find_optimal_tf} as $t_f = 394.5~(\mathrm{s})$. The path-constraint angles are selected as $\psi = 180^\circ$, $\theta = 85^\circ$, and $\phi = -30^\circ$. The control input is assumed to be unbounded.

The simulation employed the EOMs in \eqref{eq:3D_motion_eqn} together with the controller given by \eqref{eq:3D_optimal_controller}. The EOMs were integrated using MATLAB's \texttt{ode45} solver, which implements an explicit Runge-Kutta method with an adaptive time step. The controller was updated at a rate of $10~\mathrm{Hz}$. The guidance law was implemented in a closed-loop manner, except during the final second preceding each transition to a constrained arc, where it was switched to open-loop operation to ensure a smooth transition. The results were compared with those obtained using the unconstrained solution in \eqref{eq:optimal_u_unconstrained}.
%

\begin{figure}[h!]
    \centering
    \includegraphics[width=1\textwidth]{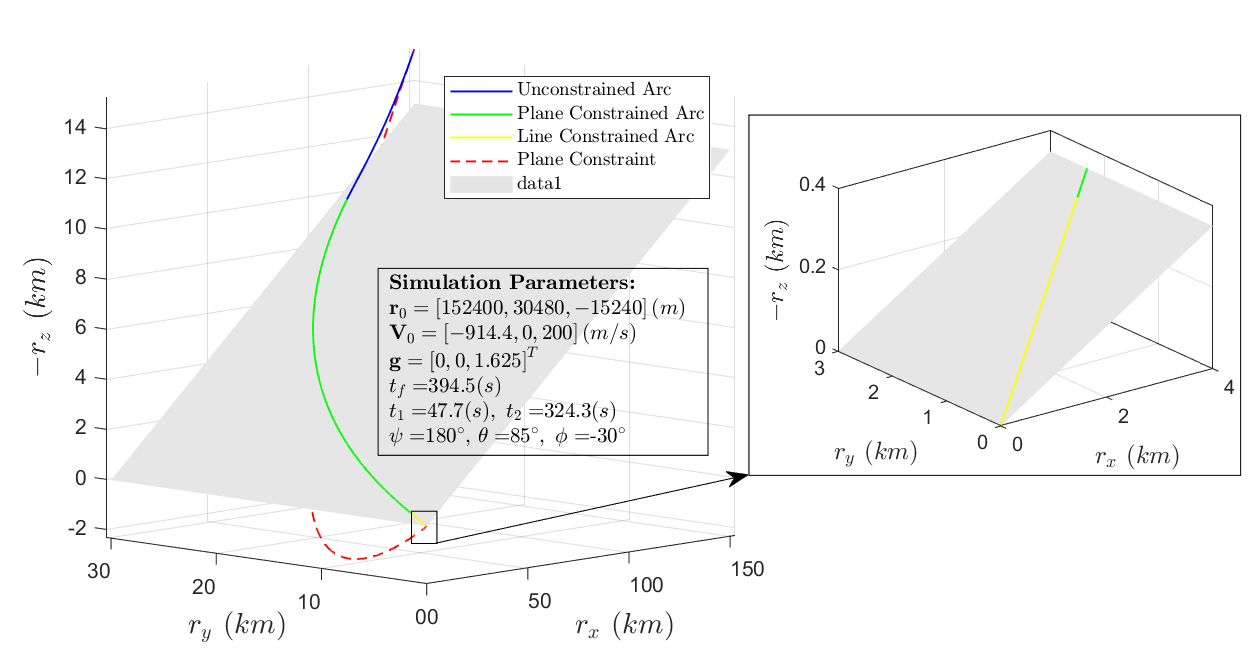}
    \caption{3D lander trajectory.}
    \label{fig:3D_trajectory}
\end{figure}

All plots are presented alongside the corresponding unconstrained solution obtained from the same initial conditions. The three-dimensional trajectory is shown in \figref{fig:3D_trajectory}, where the unconstrained, plane-constrained, and line-constrained arcs are depicted using different colors to facilitate visualization of the transitions between phases. As expected, the unconstrained trajectory intersects the ground, whereas the constrained trajectory remains collision-free.

\begin{figure}[htbp]
    \centering

    \begin{subfigure}[b]{0.47\textwidth}
        \centering
        \includegraphics[width=\textwidth]{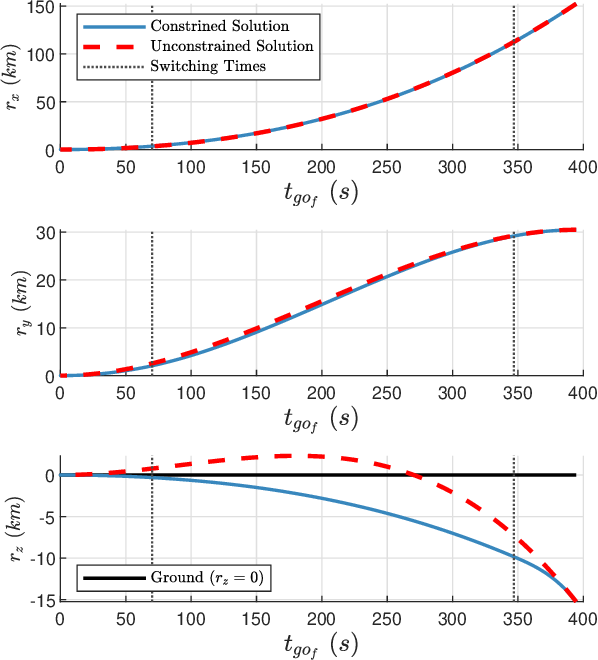}
        \caption{Position components.}
        \label{fig:xyz_position}
    \end{subfigure}
    \hfill
    \begin{subfigure}[b]{0.47\textwidth}
        \centering
        \includegraphics[width=\textwidth]{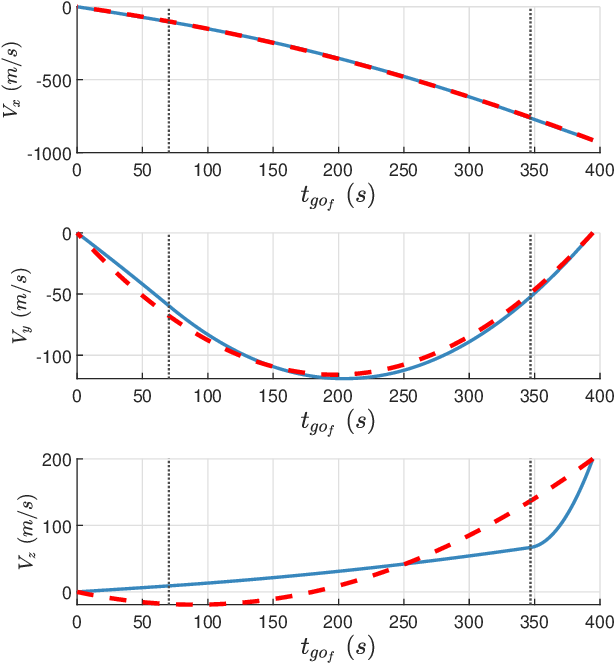}
        \caption{Velocity components.}
        \label{fig:xyz_velocity}
    \end{subfigure}

    \vspace{1em} 

    \begin{subfigure}[b]{1\textwidth}
        \centering
        \includegraphics[width=\textwidth]{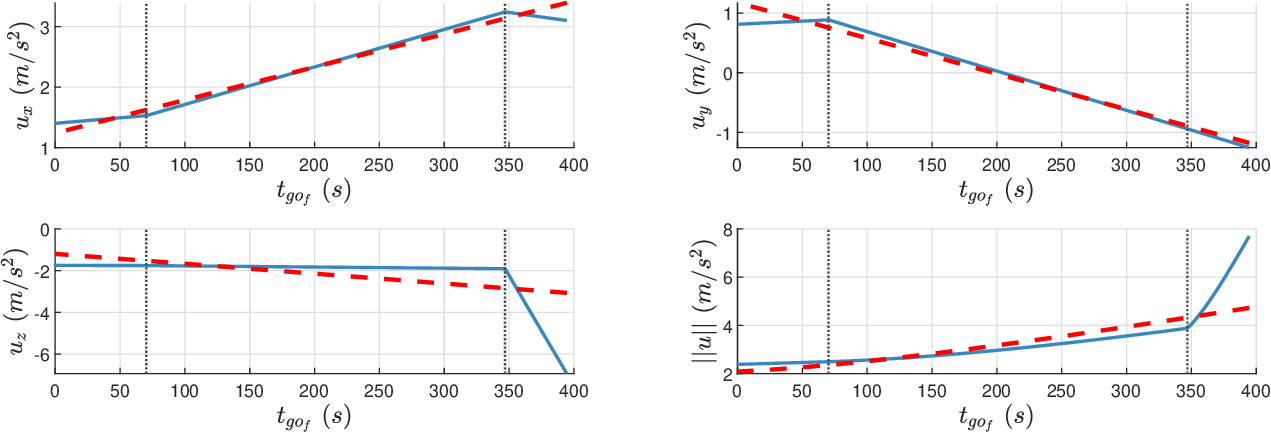}
        \caption{Control components.}
        \label{fig:xyz_control}
    \end{subfigure}
        
    \vspace{1em} 

    \begin{subfigure}[b]{0.6\textwidth}
        \centering
        \includegraphics[width=\textwidth]{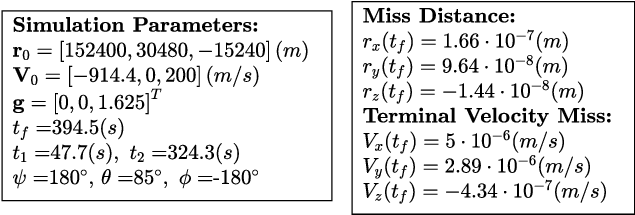}
        \caption{Simulation parameters and constrained terminal misses.}
        \label{fig:xyz_Simulation_Parameters}
    \end{subfigure}

    \caption{Simulation results in the $(x,y,z)$ frame.}
    \label{fig:3D_components_xyz}
\end{figure}

\begin{figure}[htbp]
    \centering

    \begin{subfigure}[b]{0.47\textwidth}
        \centering
        \includegraphics[width=\textwidth]{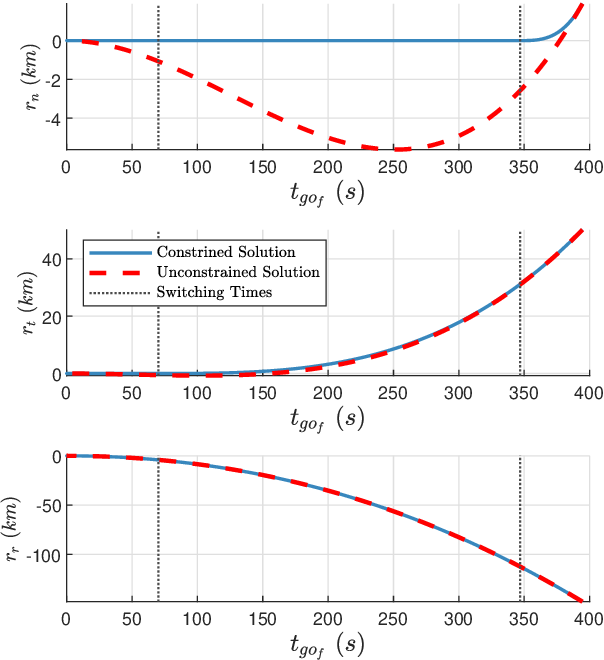}
        \caption{Position components.}
        \label{fig:ntr_position}
    \end{subfigure}
    \hfill
    \begin{subfigure}[b]{0.47\textwidth}
        \centering
        \includegraphics[width=\textwidth]{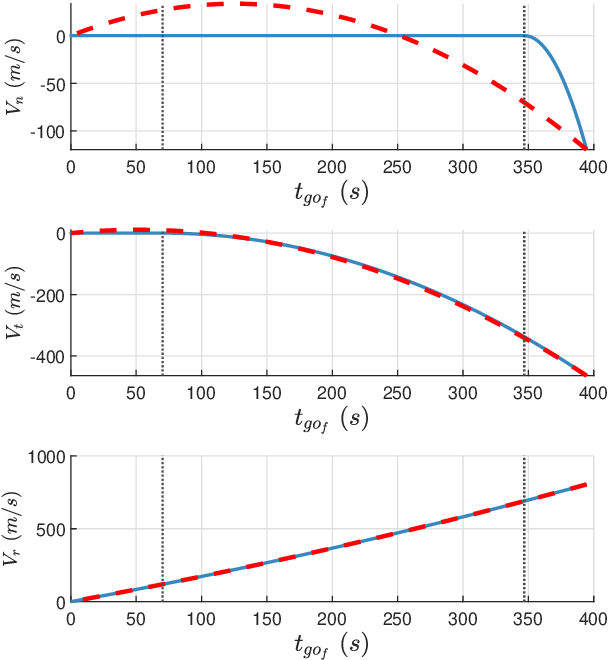}
        \caption{Velocity components.}
        \label{fig:ntr_velocity}
    \end{subfigure}

    \vspace{1em} 

    \begin{subfigure}[b]{1\textwidth}
        \centering
        \includegraphics[width=\textwidth]{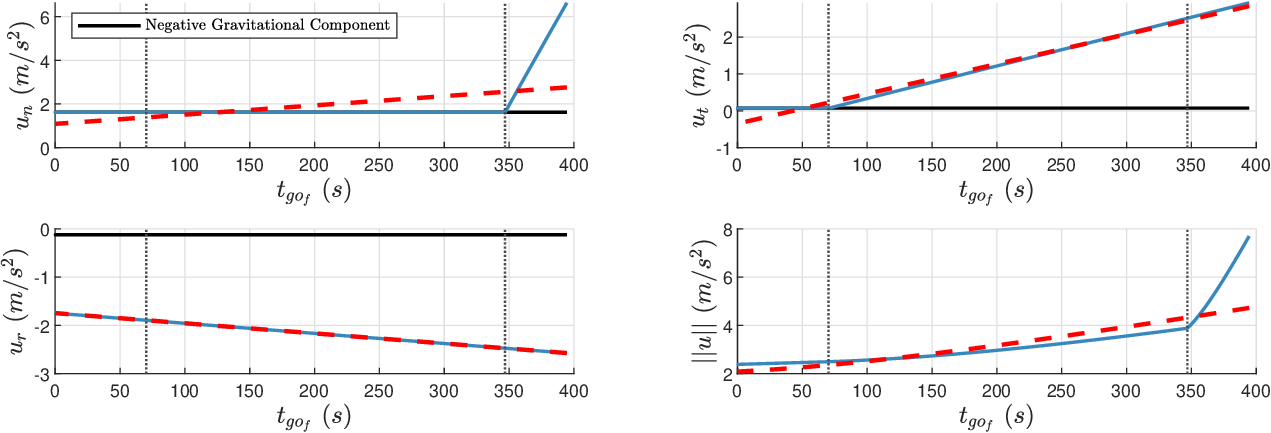}
        \caption{Control components.}
        \label{fig:ntr_control}
    \end{subfigure}
    
    \vspace{1em} 

    \begin{subfigure}[b]{0.6\textwidth}
        \centering
        \includegraphics[width=\textwidth]{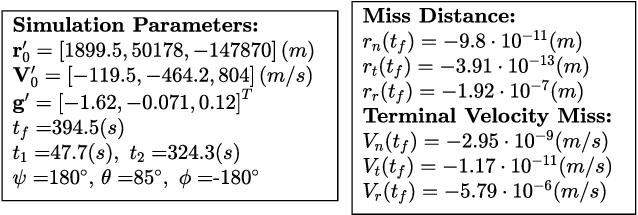}
        \caption{Simulation parameters and constrained terminal misses.}
        \label{fig:ntr_Simulation_Parameters}
    \end{subfigure}

    \caption{Simulation results in the $(n,t,r)$ frame.}
    \label{fig:3D_components_ntr}
\end{figure}

The complete simulation results, including the position, velocity, and control histories in both the $(x,y,z)$ and $(n,t,r)$ coordinate frames, are presented in \figref{fig:3D_components_xyz} and \figref{fig:3D_components_ntr}, respectively. The results demonstrate that the proposed guidance law guides the lander to the target with negligible terminal position and velocity errors while satisfying both the plane and line constraints.

Furthermore, the control profile is piecewise linear, consistent with the analytical derivation. In the $(n,t,r)$ frame, the controller cancels the gravitational components normal to the active constraints, as predicted by the analytical solution. It is also observed that the constrained solution requires a noticeably higher control effort than the unconstrained solution. This increase results from the intentionally aggressive initial velocity selected for the scenario, which was chosen to clearly illustrate the benefits of the proposed constrained guidance law.

\tableref{table:Terminal miss, velocity error and cost function} summarizes the terminal position and velocity errors, as well as the final cost, for both the proposed guidance law and the unconstrained solution. The constrained solution achieves terminal position and velocity errors that are negligible and comparable to those of the unconstrained solution, while, as expected, incurring a higher overall cost. This increase is required to satisfy the path constraints and prevent ground collision. The residual terminal errors are practically zero and arise primarily from numerical integration errors.
\begin{table}[h!]
    \centering
    \begin{tabular}{|m{5.2em}||c c c c c c c|} 
        \hline
         & $r_{x_f} (m)$ & $r_{y_f} (m)$ & $r_{z_f} (m)$ & $v_{x_f} (m/s)$ & $v_{y_f} (m/s)$ & $v_{z_f} (m/s)$ & $J (m^2/s^3)$\\ [0.5ex] 
        \hline\hline
        Constrained & $1.66 \cdot 10^{-7}$ & $9.64 \cdot 10^{-8}$ & $-1.44 \cdot 10^{-8}$ & $5 \cdot 10^{-6}$ & $2.89 \cdot 10^{-6}$ & $-4.34 \cdot 10^{-7}$ & $2.32 \cdot 10^{3}$\\ 
        Unconstrained & $7.3 \cdot 10^{-7}$ & $-1.15 \cdot 10^{-6}$ & $-8.42 \cdot 10^{-7}$ & $2.19 \cdot 10^{-5}$ & $-3.44 \cdot 10^{-5}$ & $-2.52 \cdot 10^{-5}$ & $2.18 \cdot 10^{3}$\\ 
        \hline
    \end{tabular}
    \caption{Terminal miss, velocity error, and cost function.}
    \label{table:Terminal miss, velocity error and cost function}
\end{table}

\subsection{Complete Pyramid Simulation} \label{subsec:CompletePyramidSimulation}
By combining four planar constraints, the complete square-pyramid constraint can be enforced. Simulations were performed for a range of initial positions, with nine different initial velocities considered for each position. For each case, a nominal initial velocity was defined, and perturbations of $\pm 300~(\mathrm{m/s})$ in the $x$- and $y$-velocity components were examined.

The pyramid constraint was defined with a square base. The rotation angles of the four bounding planes were selected as $\psi\in\{0^{\circ},\ 90^{\circ},\ 180^{\circ},\ 270^{\circ}\}$, with $\theta = 85^{\circ}$ for all planes. The angle $\phi$ was computed for each constraint line as described in Appendix B. All remaining simulation parameters were identical to those used in the sample run, and the final time was kept fixed across all scenarios.
For each set of initial conditions, the appropriate guidance law was selected based on the anticipated constraint activation. Evaluation of $t_1$ and $t_2$ from \eqref{eq:3D_tgo1} and \eqref{eq:tgo2}, respectively, enables determination of whether any constraint becomes active and, if so, identification of the first plane or line constraint encountered by the trajectory.

\begin{figure}
    \centering
    \includegraphics[width=0.65\linewidth]{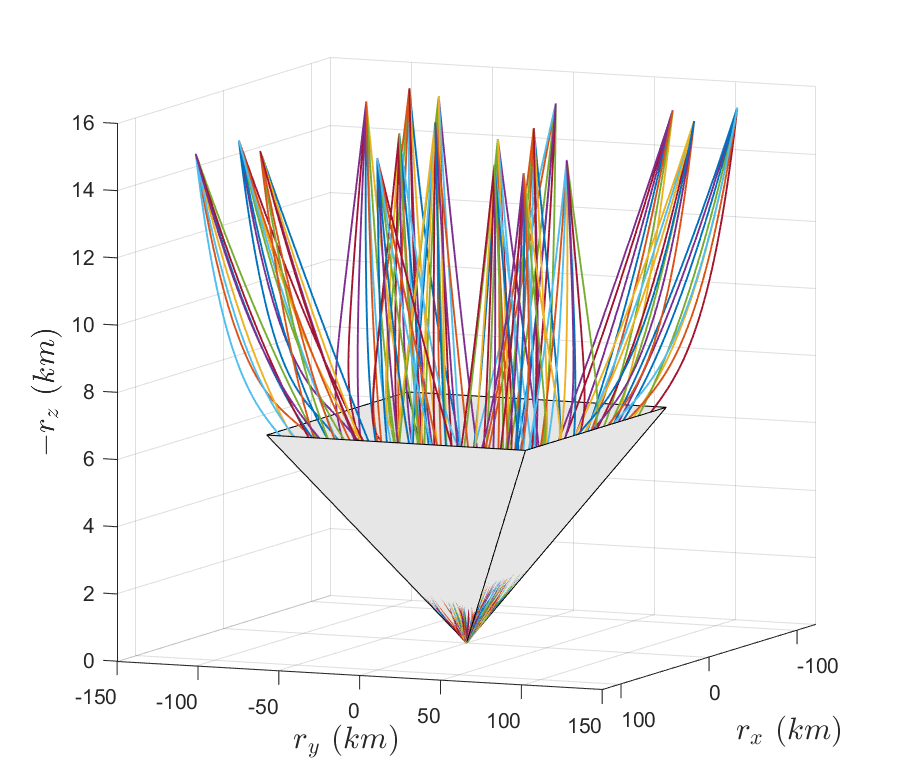}
    \caption{Complete pyramid constraint simulation for different initial conditions.}
    \label{fig:pyramid_simulation5}
\end{figure}

\begin{figure}
    \centering

    \begin{subfigure}[b]{0.48\textwidth}
        \centering
        \includegraphics[width=\textwidth]{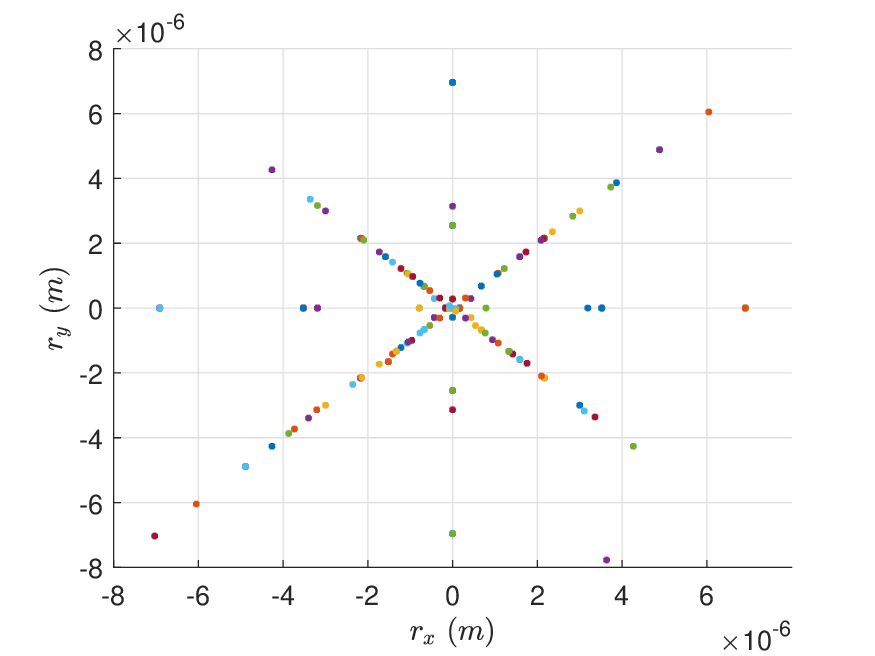}
        \caption{Position miss.}
        \label{fig:miss_distances_pyramid}
    \end{subfigure}
    \hfill
    \begin{subfigure}[b]{0.48\textwidth}
        \centering
        \includegraphics[width=\textwidth]{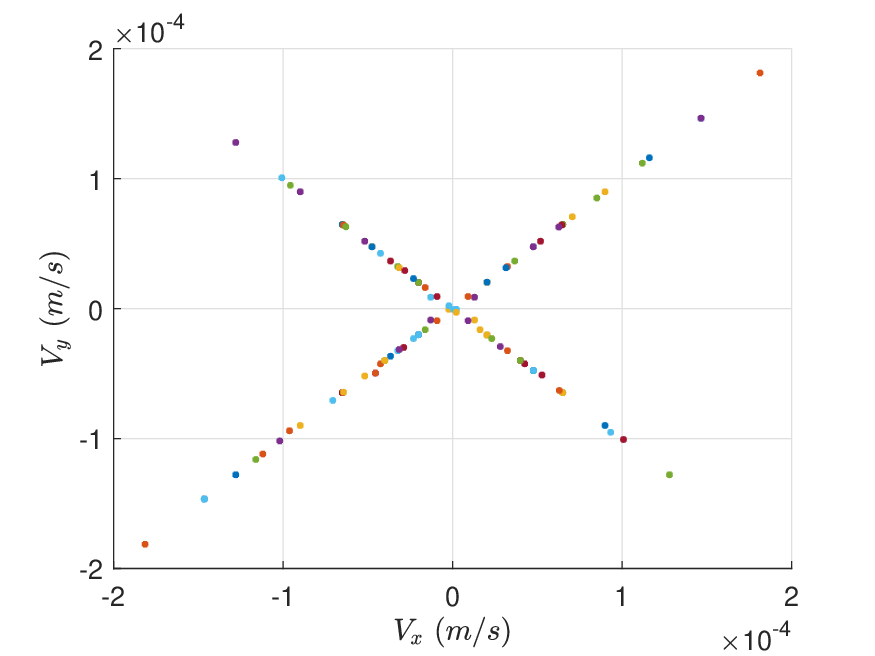}
        \caption{Velocity miss.}
        \label{fig:miss_velocity_pyramid}
    \end{subfigure}

    \caption{Terminal position and velocity misses of the complete pyramid constraint simulation.}
    \label{fig:terminal_miss}
\end{figure}

The resulting trajectories are shown in \figssref{fig:pyramid_simulation5}{fig:complete_pyramid}. As expected, none of the trajectories violate the pyramid constraint. The terminal position and velocity errors in the $x$- and $y$-directions are presented in \figref{fig:terminal_miss}. The position errors remain below $8\times10^{-6}~(\mathrm{m})$, while the velocity errors are generally below $1\times10^{-5}~(\mathrm{m/s})$ and never exceed $2\times10^{-4}~(\mathrm{m/s})$.

These results demonstrate that the proposed guidance law maintains high terminal accuracy while satisfying the square-pyramid path constraint over a broad range of initial conditions.

\begin{figure}
    \centering

    \begin{subfigure}[b]{0.49\textwidth}
        \centering
        \includegraphics[width=\textwidth]{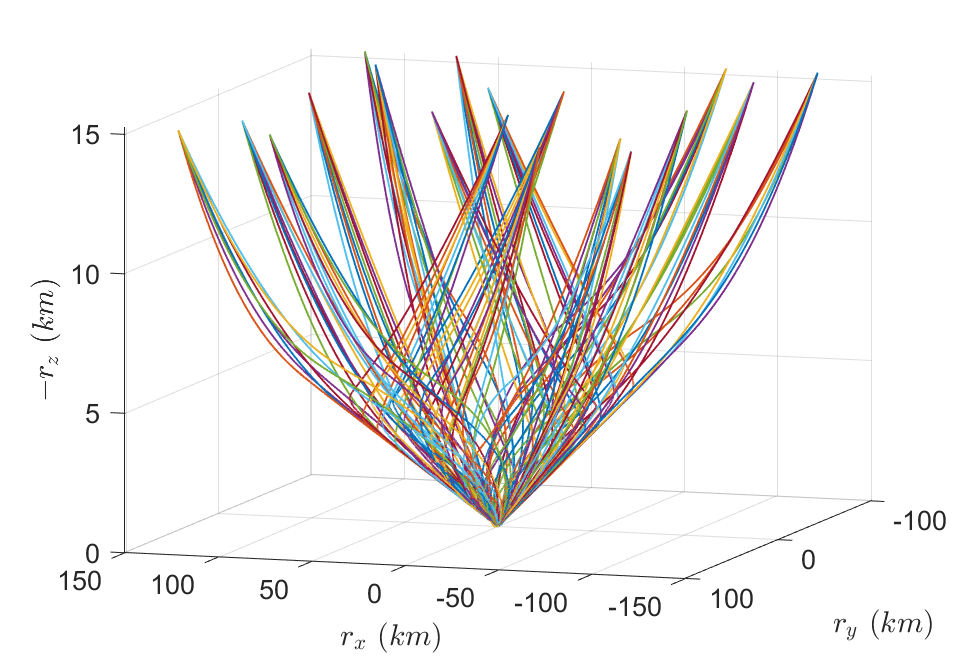}
        \caption{3D view.}
        \label{fig:pyramid_simulation}
    \end{subfigure}
    \hfill
    \begin{subfigure}[b]{0.49\textwidth}
        \centering
        \includegraphics[width=\textwidth]{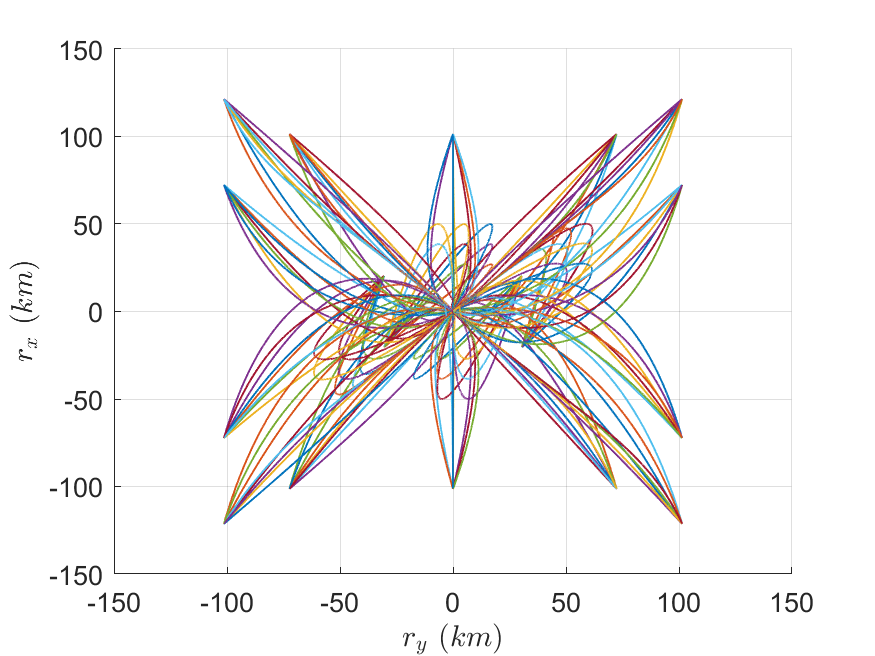}
        \caption{$x-y$ plane.}
        \label{fig:pyramid_simulation2}
    \end{subfigure}

    \vspace{1em} 

    \begin{subfigure}[b]{0.49\textwidth}
        \centering
        \includegraphics[width=\textwidth]{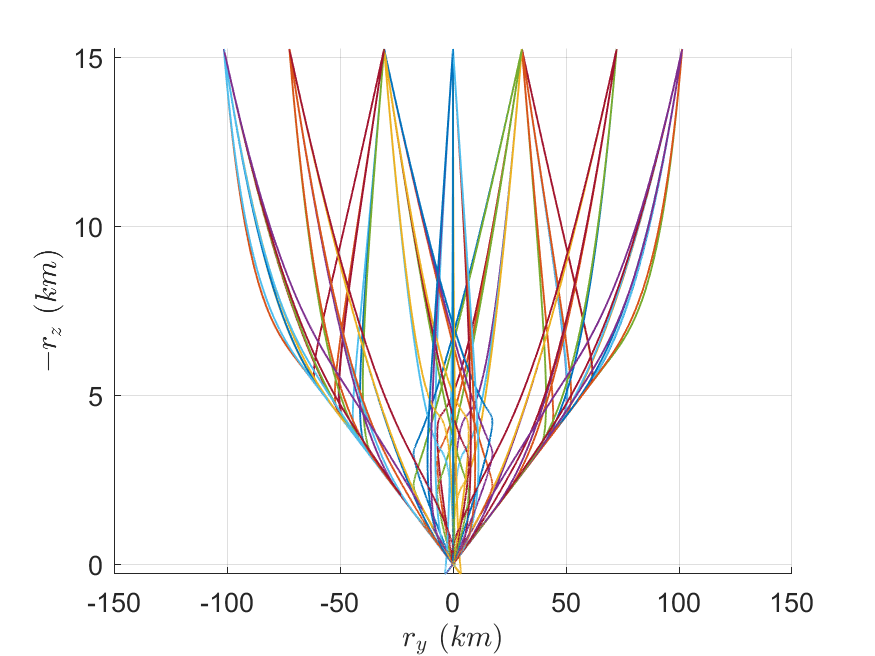}
        \caption{$y-z$ plane.}
        \label{fig:pyramid_simulation}
    \end{subfigure}
    \hfill
    \begin{subfigure}[b]{0.49\textwidth}
        \centering
        \includegraphics[width=\textwidth]{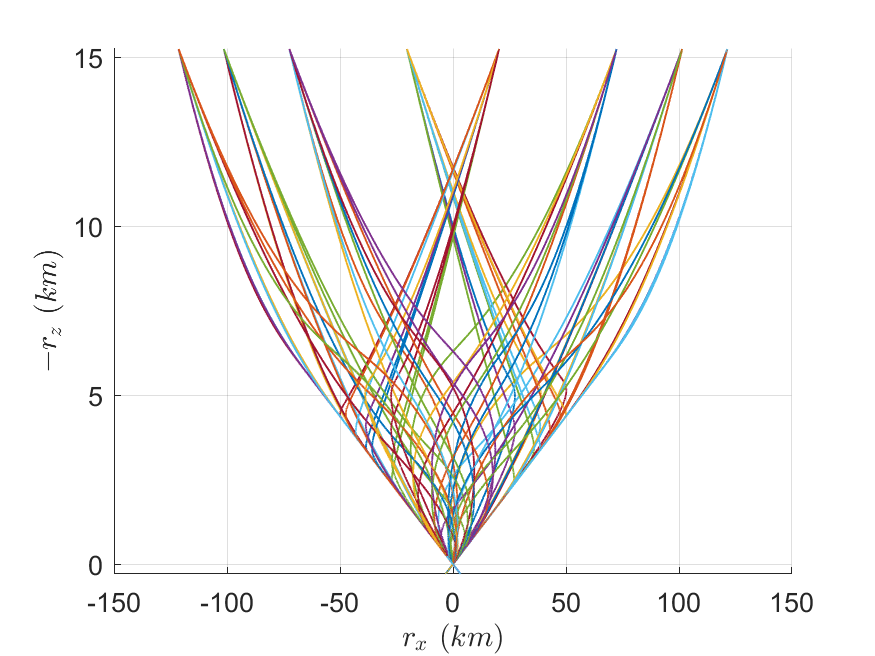}
        \caption{$x-z$ plane.}
        \label{fig:pyramid_simulation4}
    \end{subfigure}

    \caption{Different views of the complete pyramid constraint simulation for different initial conditions.}
    \label{fig:complete_pyramid}
\end{figure}

\section{Conclusions} \label{sec:Conclusions}

In this paper, an analytical optimal-control-based soft-landing guidance law with inequality path constraints was developed. The proposed guidance law simultaneously guarantees ground-collision avoidance and approach-angle control by constraining the optimal trajectory to remain within an inverted pyramid originating at the landing point. The formulation minimizes a quadratic control-effort cost while satisfying terminal position and velocity constraints.
A three-dimensional point-mass model with linear dynamics and a constant gravitational field was employed. Analytical open-loop and closed-loop solutions, together with the optimal final time, were derived using Pontryagin’s Minimum Principle and the optimality conditions at the transitions between unconstrained and constrained arcs. The resulting guidance law is continuous, piecewise linear in time, and nonlinear in the states when implemented in closed loop. When a constraint becomes active, the controller cancels the gravitational component normal to the constraint, causing the trajectory to remain on the active constraint. Furthermore, the analysis shows that each additional active constraint reduces the optimal final time. The proposed guidance law was evaluated through simulations over a range of initial conditions, demonstrating high terminal accuracy and consistent satisfaction of the path constraints.

\section*{Appendix A: Transition Matrix Derivation}

This Appendix presents the derivation of the Transition Matrix components for the kinematics used in this paper.

From the transition matrix properties 
\begin{equation}\label{Trans_matrix_prop}
    \frac{d}{dt} {\boldsymbol \Phi}(t_*,t)  = -{\boldsymbol{\Phi}} (t_*,t) {\bf A} , \quad {\boldsymbol \Phi}(t_*,t_*) = \textbf{I}
\end{equation} 
Differentiation the Transiton Matrix with respect to time-to-go, $t_{go_*}=t_*-t$,  and substituting \eqref {Trans_matrix_prop}, yields 
\begin{equation}  \label{Trans_matrix_prop_tgo_*}
    \frac{d}{dt_{{go_*}}} {\boldsymbol \Phi}(t_*,t)  = \frac{d}{dt} {\boldsymbol \Phi}(t_*,t)  \cdot \frac{dt}{dt_{{go_*}}} = - \frac{d}{dt} {\boldsymbol \Phi}(t_*,t)  = {\boldsymbol{\Phi}} (t_*,t) {\bf A}
\end{equation}
Let
\begin{equation}
    \label{solve_trans-matrix}
    {\boldsymbol \Phi}(t_*,t) = {\boldsymbol \Phi}(t_{{go_*}}) =  {\begin{bmatrix}
    [{{\boldsymbol \phi}_{rr}}]_{n \times n} & [{{\boldsymbol \phi}_{rV}}]_{n\times n} \\ 
    [{{\boldsymbol \phi}_{Vr}}]_{n\times n} & [{{\boldsymbol \phi}_{VV}}]_{n\times n}
    \end{bmatrix}}_{2n \times 2n}, \quad
     {\bf A} = \begin{bmatrix}
        {\bf{[0]}}_n & {\bf{[I]}}_n \\
        {\bf{[0]}}_n & {\bf{[0]}}_n
    \end{bmatrix}
\end{equation} 
Substituting \eqref{solve_trans-matrix} into \eqref{Trans_matrix_prop_tgo_*}, integrating with respect to $t_{go_*}$ and using the boundary conditions, yields 
\begin{subequations} \label{eq:phi1x}
\begin{align}
\label{eq:phi11}
    \frac{{d{[{{\boldsymbol \phi}_{rr}}]}}_{{n \times n}}} {dt_{{go_*}}} &= [{\bf{0}}]_{n} , & \quad [{{\boldsymbol \phi}_{rr}}](t_{{go_*}}=0) = [{\bf{I}}]_{n} & \quad  \Rightarrow [{{\boldsymbol \phi}_{rr}}](t_{{go_*}}) = [{\bf{I}}]_{n} \\ 
        \label{eq:phi12}
    \frac{{d{[{{\boldsymbol \phi}_{rV}}]}}_{{n \times n}}} {dt_{{go_*}}} &= {[{{\boldsymbol \phi}_{rr}}(t_{{go_*}})]}_{n \times n}, &\quad [{{\boldsymbol \phi}_{rV}}](t_{{go_*}}=0) = [{\bf{0}}]_{n} &\quad  \Rightarrow [{{\boldsymbol \phi}_{rV}}](t_{{go_*}}) = t_{{go_*}} \cdot  [{\bf{I}}]_{n} \\
    \label{eq:phi21}
    \frac{{d{[{{\boldsymbol \phi}_{Vr}}]}}_{{n \times n}}} {dt_{{go_*}}} &= [{\bf{0}}]_{n} , & \quad [{{\boldsymbol \phi}_{Vr}}](t_{{go_*}}=0) = [{\bf{0}}]_{n} & \quad  \Rightarrow [{{\boldsymbol \phi}_{Vr}}](t_{{go_*}}) = [{\bf{0}}]_{n} \\
    \label{eq:phi22}
    \frac{{d{[{{\boldsymbol \phi}_{VV}}]}}_{{n \times n}}} {dt_{{go_*}}} &= {[{{\boldsymbol \phi}_{Vr}}(t_{{go_*}})]}_{n \times n}, &\quad [{{\boldsymbol \phi}_{VV}}](t_{{go_*}}=0) = [{\bf{I}}]_{n} &\quad  \Rightarrow [{{\boldsymbol \phi}_{rV}}](t_{{go_*}}) = [{\bf{I}}]_{n} 
\end{align}
\end{subequations} 
The Transition Matrix is, therefore
\begin{equation} \label{transition_matrix}
    {\boldsymbol \Phi}(t_*,t) = {\boldsymbol \Phi}(t_{{go_*}}) =  {\begin{bmatrix}
    {\bf{[I]}}_n & t_{{go_*}}{\bf{[I]}}_n \\ 
    {\bf{[0]}}_n & {\bf{[I]}}_n
    \end{bmatrix}}_{2n \times 2n}
\end{equation}
\label{APPTransition}

\section*{Appendix B: Derivation of the rotation angle $\phi$ for a right square pyramid}
Our objective is to determine the rotation angle $\phi$ that aligns the vector $\hat{\mathbf{t}}$ perpendicular to the line formed by the intersection of two faces.
When the optimal trajectory transitions to one of the pyramid's faces, it can violate either the left-hand line (defined by the intersection with the left face) or the right-hand line (defined by the intersection with the right face).

\begin{figure}[htbp]
    \centering
    \begin{subfigure}[b]{0.275\textwidth}
        \centering
        \includegraphics[width=\textwidth]{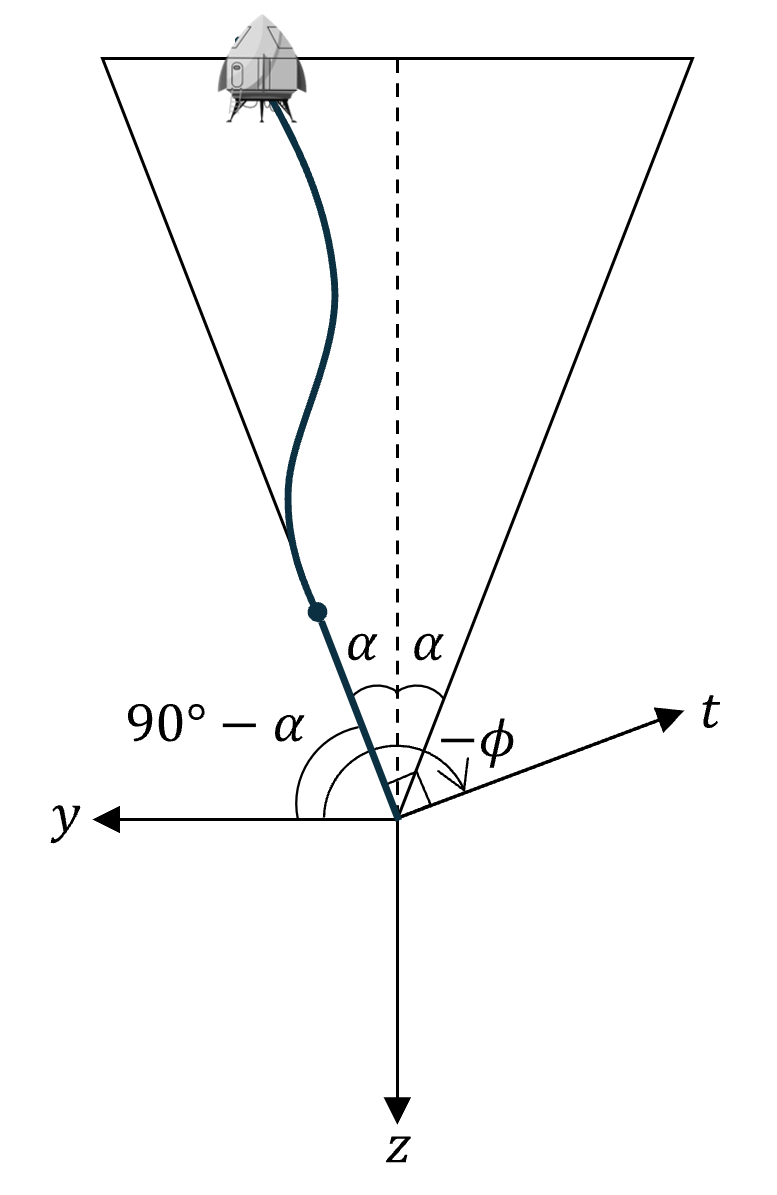}
        \caption{Left line violation.}
        \label{fig:left_line}
    \end{subfigure}
    \hfill
    \begin{subfigure}[b]{0.245\textwidth}
        \centering
        \includegraphics[width=\textwidth]{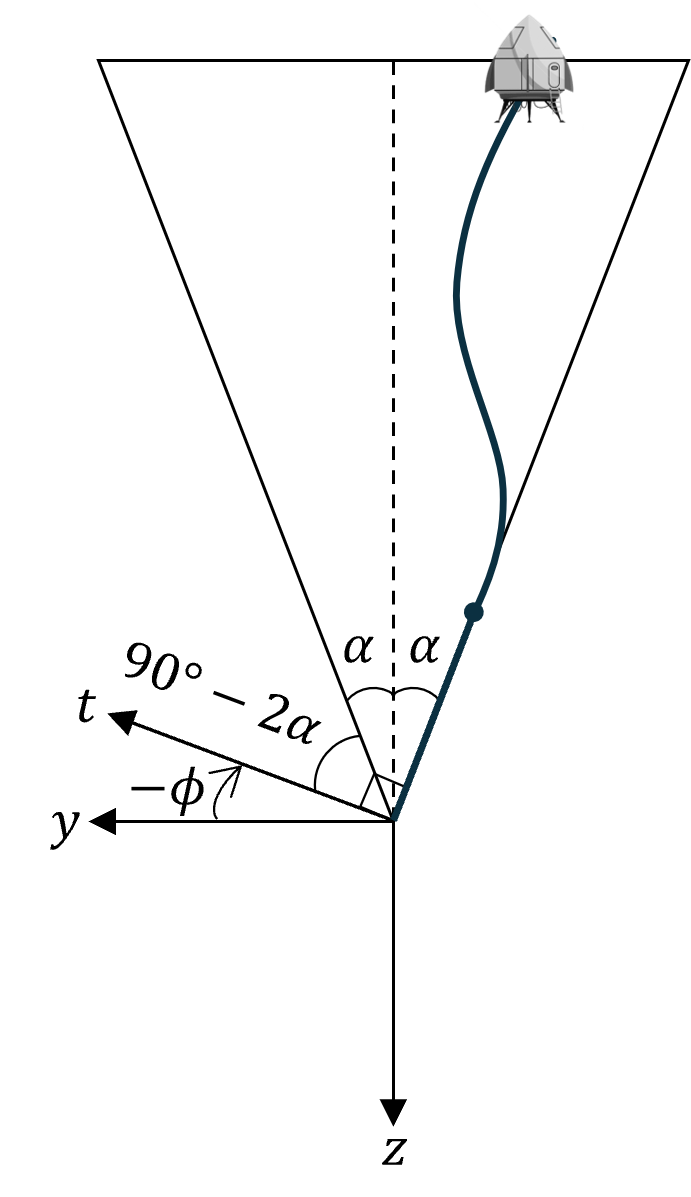}
        \caption{Right line violation.}
        \label{fig:right_line}
    \end{subfigure}
    \hfill
    \begin{subfigure}[b]{0.35\textwidth}
        \centering
        \includegraphics[width=\textwidth]{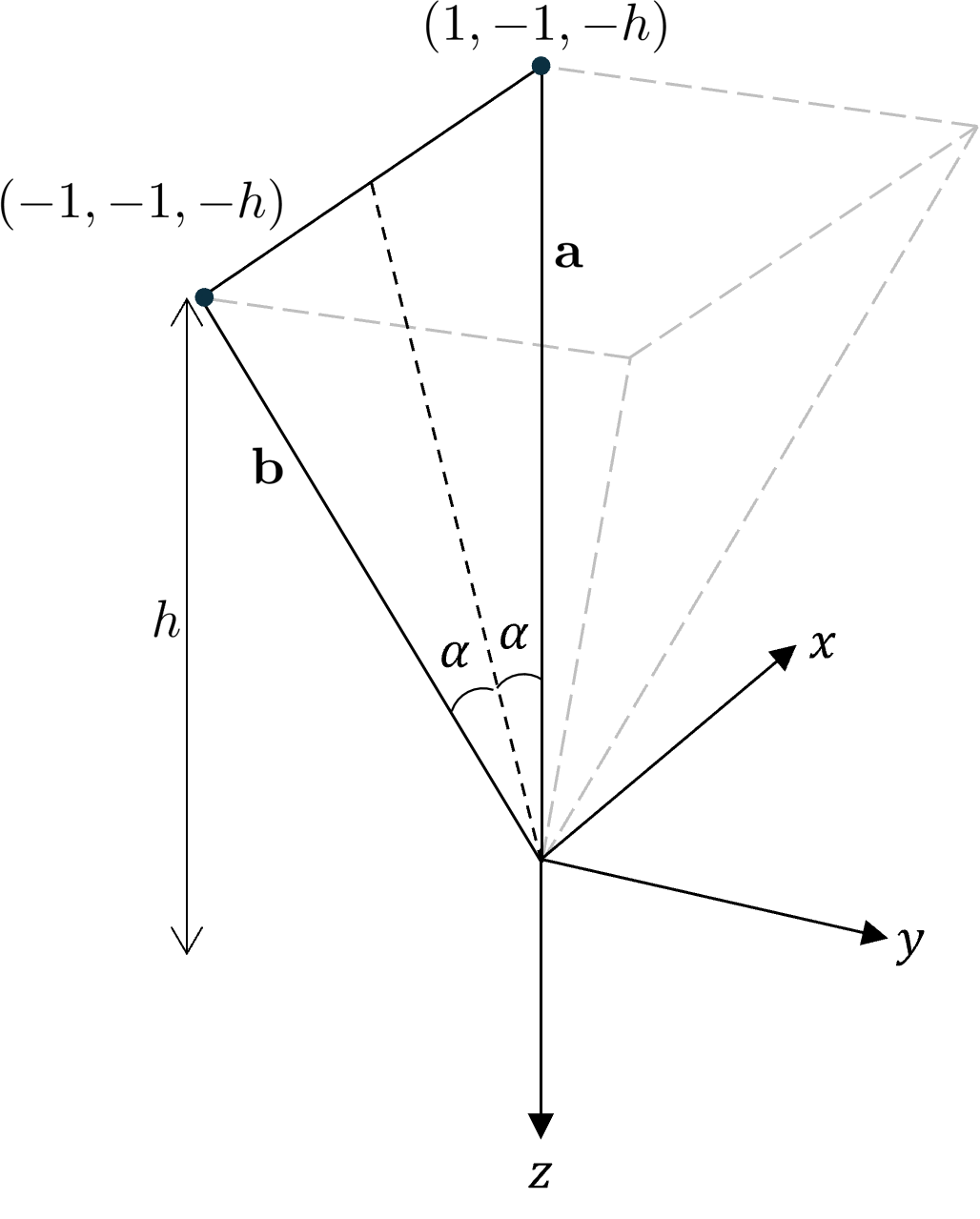}
    \caption{Calculating $\alpha$.}
    \label{fig:pyramid_for_alpha}
    \end{subfigure}

    \caption{Angles description.}
    \label{fig:trangle}
\end{figure}

\figref{fig:trangle} describes the rotation angle $\phi$, and an additional angle $\alpha$, for a left side violation and a right side violation.
\figssref{fig:left_line}{fig:right_line} describe a specific plane of the pyramid, after two rotations of the $(x, y, z)$ frame.
The relation between $\phi$ and $\alpha$ is as follows, according to \figssref{fig:left_line}{fig:right_line}
\begin{equation} \label{eq:phi_alpha}
   - \phi = \begin{cases}
        \alpha, & \text{Right constraint violation} \\
        \pi - \alpha, & \text{Left constraint violation} \\
    \end{cases}
\end{equation}

To determine $\alpha$, we look at \figref{fig:pyramid_for_alpha} that describes a section of the pyramid face. 
Because the pyramid is a right square let
\begin{equation}
    {\bf{a}} = \begin{bmatrix}
        1 & -1 & -h
    \end{bmatrix}^T ,\quad {\bf{b}} = \begin{bmatrix}
        -1 & -1 & -h
    \end{bmatrix}^T
\end{equation}
The dot product of ${\bf{a}}$ and ${\bf{b}}$ yields
\begin{equation} \label{eq:alpha}
    {\bf{a}}\cdot {\bf{b}} = \vert\vert{\bf{a}}\vert\vert\cdot\vert\vert{\bf{b}}\vert\vert\cdot\cos{\left(2\alpha\right)} = h^2 \quad\Rightarrow\quad \alpha = \frac{1}{2}\cos^{-1}{\left( \frac{h^2}{\vert\vert{\bf{a}}\vert\vert\cdot\vert\vert{\bf{b}}\vert\vert} \right)}
\end{equation}
Because ${\bf{b}}$ lies on the face (in the plane defined by ${\bf{a}}$ and ${\bf{b}}$)
\begin{equation}
    b_n={\begin{bmatrix}
    1 & 0  & 0
    \end{bmatrix}}
    {\bf D}_{nx}{\bf{b}}=-\cos\theta\cos\psi -\cos\theta\sin\psi +h\sin\theta=0
\end{equation}
Rearranging and substituting $\psi=\pi/2$, we get
\begin{equation}
    h = \left|\frac{-\cos{(\theta)}\cos{(\psi)} - \cos{(\theta)}\sin{(\psi)}}{-\sin{(\theta)}}\right| = \left|\frac{\cos{(\theta)}}{\sin{(\theta)}}\right|
\end{equation}
which can be used to calculate $\phi$ by substituting into \eqref{eq:alpha} and \eqref{eq:phi_alpha}.
\label{APPPhiDerivation}

\section*{Appendix C: The Unconstrained Solution}
This Appendix presents the optimal solution for the unconstrained case, i.e., when the path constraints remain inactive throughout the trajectory. For consistency with the main text, the derivation is presented in the $(n,t,r)$ frame. However, because the coordinate transformation does not alter the dynamics, the expressions in \eqref{eq:3D_lambda} and \eqref{eq:3D_u_unconstrained}, and consequently the resulting solution, are identical in the $(x,y,z)$ frame. Although these results were originally derived in the $(x,y,z)$ frame in \cite{1997}, they are reproduced here in the $(n,t,r)$ frame for completeness.

Using the state-transition matrix together with the system dynamics and terminal conditions, we obtain
\begin{equation}
    {\bf{x}}'_f  = \boldsymbol{\Phi}(t_{go_f}) {\bf{x}}'(t) + \int_{t}^{t_f}\boldsymbol{\Phi}(t_f-\tau){\bf{B}u}'^*(\tau)d\tau + \int_{t}^{t_f}\boldsymbol{\Phi}(t_f-\tau){\bf{C}{g}}'d\tau = {\bf{0}}
\end{equation}
Substituting $t_{go_f}=t_{go_1}$ into \eqref{eq:3D_lambda} and \eqref{eq:3D_u_unconstrained}, integrating, and rearranging yields
\begin{equation}  \label{eq:L_i}
    \begin{bmatrix}
        L_1 &
        L_2 &
        L_3 &
        L_4 &
        L_5 &
        L_6
    \end{bmatrix}^T = \frac{12}{t_{go_f}^4} \begin{bmatrix}
        t_{go_f}{\bf{r}}'(t) + \dfrac{t_{go_f}^2}{2}{\bf{V}}'(t)\\
        -\dfrac{t_{go_f}^2}{2}{\bf{r}}'(t) - \dfrac{t_{go_f}^3}{6}{\bf{V}}'(t) + \dfrac{t_{go_f}^4}{12}{\bf{g}}'
    \end{bmatrix}
\end{equation}
Substituting \eqref{eq:L_i} into \eqref{eq:3D_u_unconstrained} yields the following optimal controller
\begin{equation} \label{eq:optimal_u_unconstrained}
    {\bf u}'^* = -\frac{6{\bf{r}}'(t)}{t_{go_f}^2} - \frac{4{\bf{V}}'(t)}{t_{go_f}} - {\bf{g}}'
\end{equation}

Following similar steps to \secref{subsec:OptimalFinalTime}, we can obtain an equation for the optimal final time in the unconstrained case
\begin{equation} \label{eq:find_optimal_tf_unconstrained}
    {{\Vert {\bf{g}}'\Vert }^2}\left(t_{f}^*\right)^4 - 4{{\Vert {\bf{V}}'(t_0)\Vert }^2}\left(t_{f}^*\right)^2 - 24{\bf{r}}'^T(t_0){\bf{V}}'(t_0)t_{f}^* - 36{{\Vert {\bf{r}}'(t_0)\Vert }^2} = 0
\end{equation}
The optimal final time for the unconstrained case is the minimum real positive solution of \eqref{eq:find_optimal_tf_unconstrained} \cite{ZEMZEV}. 

\label{APPNoLineViolationSolution}

\section*{Appendix D: Solution When Only the Plane Constraint Is Active}
This Appendix presents the optimal solution for a transition from an unconstrained arc to a constrained arc on a plane without a transition to a line constraint (i.e., the trajectory remains inside the triangular face and does not transition to an edge of the triangle).

\subsection{Plane-Constrained Arc and Switching Point Conditions}

Substituting $t_{go_f} = t_{go_2}$ into \eqref{eq:3D_mu} and \eqref{eq:3D_u_constrained} yields
\begin{equation} \label{eq:3D_mu_1point}
  \boldsymbol{\mu}(t_{go_f}) = \begin{bmatrix}
       M_1 & M_2 & (M_1\cdot t_{go_f} + M_3) & (M_2\cdot t_{go_f} + M_4) 
   \end{bmatrix}^T, \quad
    \bar{{\bf{u}}}'^*  = -\begin{bmatrix}
        \mu_3 & \mu_4 \end{bmatrix}^T
\end{equation} 
Using the state-transition matrix together with the system dynamics and terminal conditions, we obtain
\begin{equation}
    {\bf\bar{{\bf{x}}}'}_f = \bar{\boldsymbol{\Phi}}'(t_{go_f}) {\bf\bar{{\bf{x}}}'}(t) + \int_{t}^{t_f}{\bar{\boldsymbol{\Phi}}'(t_f-\tau)\bar{\bf{B}}\bar{\bf u}'^*(\tau)d\tau} + \int_{t}^{t_f}{\bar{\boldsymbol{\Phi}}'(t_f-\tau)\bar{\bf{C}}{\bf{g}'}d\tau}= {\bf{0}}
\end{equation}
Substituting \eqref{eq:3D_mu_1point}, integrating, and rearranging yields
\begin{equation}  \label{eq:3D_mu_APP_C} 
    \begin{bmatrix}
        M_1 &
        M_2 &
        M_3 &
        M_4 
    \end{bmatrix}^T = \frac{12}{t_{go_f}^4} \begin{bmatrix}
        t_{go_f}{\bf\bar{{\bf{r}}}'}(t) + \dfrac{t_{go_f}^2}{2}{\bf\bar{{\bf{V}}}'}(t)\\
        -\dfrac{t_{go_f}^2}{2}{\bf\bar{{\bf{r}}}'}(t) - \dfrac{t_{go_f}^3}{6}{\bf\bar{{\bf{V}}}'}(t) + \dfrac{t_{go_f}^4}{12}{\bf\bar{{\bf{g}}}'}
    \end{bmatrix}
\end{equation}
Similarly to \eqref{eq:3D_junc_cond}, the following conditions must be satisfied at the switching point $t_1$
\begin{subequations} \label{eq:3D_junc_cond_APP_C}
    \begin{align}
    \boldsymbol{\mu}^T(t_1^+) &= \boldsymbol{\lambda}^T(t_1^-)\bf{E} \label{eq:3D_junc_cond_1_APP_c}  \\
    \mathcal{G}(t_1^+) &= \mathcal{H}(t_1^-) + \boldsymbol{\lambda}^T(t_1^-) \left( {\bf{W}}{\bf{y}}_t + {\bf{E}}{\bar{{\bf{x}}}}'_t \right) \label{eq:3D_junc_cond_2_APP_C}
    \end{align}
\end{subequations}
where $\bf{W}$ and $\bf{E}$  are given in \eqref{eq:MN_PQ} and ${\bf{y}}_t$ and ${\bar{{\bf{x}}}}'_t$ are given in \eqref{eq:Time_partial_derivs}.

Substituting \eqref{eq:3D_lambda} and  \eqref{eq:3D_mu_1point}
into the switching point condition in \eqref{eq:3D_junc_cond_1_APP_c} yields
\begin{equation} \label{eq:3D_costates_cont_APP_C}
    \begin{cases}
        \mu_1(t_1) = \lambda_2(t_1) = L_2 = M_1 \\
        \mu_2(t_1) = \lambda_3(t_1) = L_3 = M_2 \\
        \mu_3(t_1) = \lambda_5(t_1) = L_5 = M_1 \Delta t_{1f} + M_3 \\
        \mu_4(t_1) = \lambda_6(t_1) = L_6 = M_2 \Delta t_{1f} + M_4
    \end{cases} 
\end{equation}
where $\Delta t_{1f}=t_f-t_1$.
Using the switching point condition in \eqref{eq:3D_junc_cond_2_APP_C}, and following the same steps as in \secref{subsec:First_Switching_Point} we can obtain
the same results for $L_1, L_4$, and $t_{go_1}$
\begin{equation} \label{eq:3D_L4_L1_tgo1_AppC}
    L_1 = \frac{2V_n(t)}{t_{go_1}^2},  \quad L_4 = g_n, \quad t_{go_1} = -\frac{3r_n(t)}{V_n(t)}
\end{equation}

\subsection{The Combined Guidance Law}

Combining the controllers for the two phases 
\begin{equation} \label{eq:3D_optimal_controller_temp_APP_C} 
    {\bf{u}'}^* = \begin{cases}
        -\begin{bmatrix}
             L_1\cdot t_{go_1} + L_4 & L_2\cdot t_{go_1} + L_5 & L_3\cdot t_{go_1} + L_6
        \end{bmatrix}^T \quad ,t\leq t_1 \\
        -\begin{bmatrix}
             g_n &
             M_1\cdot t_{go_f} + M_3 & M_2\cdot t_{go_f} + M_4
        \end{bmatrix}^T \quad ,t_1<t \leq t_f
    \end{cases}
\end{equation}
and substituting Eqs. \ref{eq:3D_mu_APP_C}, (\ref{eq:3D_costates_cont_APP_C}), and \eqref{eq:3D_L4_L1_tgo1_AppC}, we can write the optimal controller as a function of the current state and the plane-constrained reduced state
\begin{equation} \label{eq:3D_optimal_controller_APP_C} 
    {\bf{u}}'^* = -\begin{bmatrix}
        g_n -\dfrac{2V_n^2(t)}{3r_n(t)}\cdot  \mathbbm{1}(t_1-t) \\
        g_t + \dfrac{6r_t(t)}{t_{go_f}^2} + \dfrac{4V_t(t)}{t_{go_f}} \\
        g_r + \dfrac{6r_r(t)}{t_{go_f}^2} + \dfrac{4V_r(t)}{t_{go_f}}
    \end{bmatrix}, \quad \mathbbm{1}(\tau) = \begin{cases}
        0 ,& \tau < 0 \\
        1 ,& \tau \geq 0
    \end{cases}
\end{equation}

Following the same steps as in \secref{subsec:OptimalFinalTime}, we can obtain the equation for the optimal final time in the case where only the plane constraint becomes active
\begin{equation} \label{eq:find_optimal_tf_plane}
    {{\Vert {\bf g}'\Vert }^2}\left(\bar{t}_{f}^{\ *}\right)^4 - 4{{\Vert \bar{\bf{V}}'(t_0)\Vert }^2}\left(\bar{t}_{f}^{\ *}\right)^2 - 24{\bar{\bf{r}}}'^T(t_0){\bar{\bf{V}}}'(t_0)\bar{t}_{f}^{\ *} - 36{{\Vert \bar{\bf{r}}'(t_0)\Vert }^2} = 0
\end{equation}
Where $\bar{t}_{f}^{\ *}$ is the optimal time in the case in which only the plane constraint becomes active. 

\label{APPNoLineViolationSolution}

\section*{Appendix E: Effect of Constraint Activation on the Optimal Final Time}
We now prove that the activation of the constraints reduces the optimal final time. To this end, consider \eqsref{eq:find_optimal_tf}, \eqref{eq:find_optimal_tf_plane}, and \eqref{eq:find_optimal_tf_unconstrained}  corresponding to the line-constrained, plane-constrained, and unconstrained cases, respectively. Define the functions $\bar{\bar{f}}$, $\bar{f}$, and $f$ such that their roots correspond to the optimal final times in the respective cases.
\begin{subequations}
    \begin{align}
      \bar{\bar{f}}\left(t_f\right) &={{\Vert {\bf g}'\Vert }^2} t_f^4 - 4V_r^2(t_0) t_f ^2 - 24r_r(t_0)V_r(t_0) t_f - 36r_r^2(t_0) = 0, \quad \bar{\bar{f}}\left(\bar{\bar{t}}_f^{\ *}\right)=0 \\ 
        {\bar{f}}\left(t_f\right) &={{\Vert {\bf g}'\Vert }^2} t_f^4 - 4{{\Vert \bar{\bf{V}}'(t_0)\Vert }^2} t_f^2 - 24{\bar{\bf{r}}}'^T(t_0){\bar{\bf{V}}}'(t_0) t_f - 36{{\Vert \bar{\bf{r}}'(t_0)\Vert }^2}, \quad  {\bar{f}}\left({\bar{t}}_f^{\ *}\right) = 0\\
        f\left(t_f\right) &={{\Vert {\bf{g}}'\Vert }^2} t_f^4 - 4{{\Vert {\bf{V}}'(t_0)\Vert }^2} t_f^2 - 24{\bf{r}}'^T(t_0){\bf{V}}'(t_0) t_f - 36{{\Vert {\bf{r}}'(t_0)\Vert }^2}, \quad f\left(t_f^{\ *}\right)=0
    \end{align}
\end{subequations}
These equations can be rewritten as
\begin{subequations} \label{eq:all_optimal_time_equations}
    \begin{align}
         \bar{\bar{f}}\left(t_f\right) &= \Vert{\bf{g}}\Vert^2 t_f^4 - 4\Vert V_r(t_0) t_f + 3r_r(t_0)\Vert^2, \quad \bar{\bar{f}}\left(\bar{\bar{t}}_f^{\ *}\right)=0 \\
        {\bar{f}}\left(t_f\right) &= 
        \bar{\bar{f}}\left(t_f\right) - 4\Vert V_t(t_0) t_f + 3r_t(t_0)\Vert^2 , \quad  {\bar{f}}\left({\bar{t}}_f^{\ *}\right) = 0 \\
        f\left(t_f\right) &= 
        {\bar{f}}\left(t_f\right)- 4\Vert V_n(t_0) t_f + 3r_n(t_0)\Vert^2, \quad f\left(t_f^{\ *}\right)=0
    \end{align}
\end{subequations}
The functions $\bar{\bar{f}}$, ${\bar{f}}$, and $f$ are continuous functions of $t_f$ and it is clear that
\begin{equation}
    \bar{\bar{f}}\left(t_f \rightarrow 0\right)\leq 0 , \quad {\bar{f}}\left(t_f \rightarrow 0\right)\leq 0 , \quad f\left(t_f \rightarrow 0\right) \leq 0
\end{equation}
In addition, by substituting the root we can see from \eqref{eq:all_optimal_time_equations} that
\begin{subequations} \label{eq:optimal_times_with_roots}
    \begin{align}
 {\bar{f}}\left(t_f^{\ *}\right) & = 4\Vert V_n(t_0) t_f + 3r_n(t_0)\Vert^2 \geq 0 \\ 
 \bar{\bar{f}}\left({\bar{t}}_f^{\ *}\right) & =  4\Vert V_t(t_0) {\bar{t}}_f^{\ *} + 3r_t(t_0)\Vert^2 \geq 0
 \end{align}
 \end{subequations} 
Therefore, because the optimal final times are the minimum real roots of \eqref{eq:all_optimal_time_equations} \cite{ZEMZEV}, we can conclude, according to the Intermediate Value Theorem, that 
\begin{subequations} \label{eq:optimal_times_relations}
    \begin{align}
 {\bar{f}}\left(t_f \rightarrow 0\right)\leq 0 , \quad
 {\bar{f}}\left({\bar{t}}_f^{\ *}\right)=0
 \quad{\bar{f}}\left(t_f^{\ *}\right) & \geq 0  \quad \Rightarrow\quad 0 \leq{\bar{t}}_f^{\ *} \leq t_f^{\ *}\\ 
\bar{\bar{f}}\left(t_f \rightarrow 0\right)\leq 0, \quad
\bar{\bar{f}}\left(\bar{\bar{t}}_f^{\ *}\right)=0, \quad
\bar{\bar{f}}\left({\bar{t}}_f^{\ *}\right) & \geq 0 \quad 
\Rightarrow\quad 0 \leq \bar{\bar{t}}_f^{\ *} \leq {\bar{t}}_f^{\ *}
 \end{align}
 \end{subequations} 
Which proves the claim that 
\begin{equation}
    0 \leq \bar{\bar{t}}_f^{\ *} \leq \bar{t}_f^{\ *} \leq t_f^{\ *}
\end{equation}

\label{APPAdditionalOptimalFinalTimeDerivations.tex
}

\section*{Acknowledgment}
The authors acknowledge the use of ChatGPT and Grammarly for editorial assistance during manuscript preparation, including grammar correction and language refinement.

\bibliography{sample}

\end{document}